\documentclass{ws-ijmpa}
\usepackage[super,compress]{cite}
\usepackage{graphicx}
\usepackage{subfigure}
\usepackage{varioref,exscale,latexsym,amsmath,amssymb}
\usepackage{bm}
\usepackage[vcentermath]{youngtab}
\usepackage[colorlinks=true, pdfstartview=FitV, linkcolor=black, citecolor=black, urlcolor=black]{hyperref}

\makeatletter
\newcommand{\xRightarrow}[2][]{\ext@arrow 0359\Rightarrowfill@{#1}{#2}}
\makeatother

\begin{document}
\markboth{Aydemir, Minic, Sun, Takeuchi}{Pati-Salam Unification from Non-commutative Geometry and the TeV-scale $W_R$ boson}

%
\catchline{}{}{}{}{}
%

\title{Pati-Salam Unification from Non-commutative Geometry and the TeV-scale $W_R$ boson}

\author{Ufuk Aydemir}
\address{Department of Physics and Astronomy, Uppsala University, Uppsala 751 20, Sweden\\
ufuk.aydemir@physics.uu.se
}

\author{Djordje Minic, Chen Sun, and Tatsu Takeuchi}
\address{Center for Neutrino Physics, Department of Physics, Virginia Tech, Blacksburg, VA 24061 U.S.A.\\
dminic@vt.edu, sunchen@vt.edu, takeuchi@vt.edu}

\maketitle

\begin{history}
\end{history}

\begin{abstract}
We analyze the compatibility of the unified left-right symmetric Pati-Salam models motivated
by non-commutative geometry and the TeV scale right-handed $W$ boson suggested by recent
LHC data.  We find that the unification/matching conditions place conflicting demands on the
symmetry breaking scales and that generating the required $W_R$ mass and coupling is non-trivial.
\end{abstract}

\ccode{12.60.-i,12.90.+b}

\section{Introduction and Overview}

Canonical discussions on possible new physics beyond the Standard Model (SM)
have been centered around the hierarchy problem and the unifications of couplings.
The current favorites among various approaches to stabilizing the low Higgs mass ($126$ GeV as found at the LHC\cite{Aad:2012tfa,Chatrchyan:2012xdj}) are supersymmetry, technicolor, and extra dimensions. 
These approaches also incorporate the philosophy of coupling unification
in Grand Unified Theories (GUT's).

To this list, we seek to add another contender, namely models based on 
the non-commutative geometry (NCG) of Connes.\cite{connes-book1,connes-book2}
In a series of papers starting from 1990,\cite{Connes:1990qp,Chamseddine:1991qh,Chamseddine:1996zu,Connes:1996gi,Chamseddine:2006ep,Connes:2006qv,Chamseddine:2007hz,Chamseddine:2010ud,Chamseddine:2012sw,Chamseddine:2013rta} 
Connes and collaborators have argued that the SM action could be derived from a particular NCG
via the construction of what they call the ``spectral'' action, \cite{Chamseddine:1996zu}
in essence geometrizing the SM and placing it on a similar footing to gravity. 
Several of the predictions that result from the approach, according to our current understanding, 
are quite remarkable:\footnote{%
We are currently working on a review article explaining how these predictions come about. \cite{review}
}
\begin{itemize}
\item The $SU(2)_L$ gauge bosons and the Higgs doublet are unified into a single ``superconnection,''
one of the consequences being that the $SU(2)_L$ gauge coupling $g_L$ and the Yukawa couplings are related
in a particular way. \cite{Connes:1990qp, Connes:1996gi, Chamseddine:2006ep, Connes:2006qv}
\item The $SU(2)_L\times U(1)_Y\times SU(3)_C$ gauge couplings satisfy an $SO(10)$ GUT-like relation, even though
the particle content of the model is that of the SM. \cite{Chamseddine:1996zu, Chamseddine:2010ud}
\item Anomaly cancellation requires the presence of both electroweak and QCD sectors, another GUT-like feature. \cite{Connes:1996gi}
\item The smallness of the Higgs boson mass can potentially be explained via an extra-dimension-like mechanism
involving a `warp'-factor.\cite{Chamseddine:2010ud}
\end{itemize}
The approach, of course, is not without its problems:
\begin{itemize}
\item The GUT-like relations on the gauge couplings can only be imposed at a single scale,
so one must interpret the NCG spectral action as that which `emerges' from an underlying NCG theory at
the `unification' scale.  
\item Quantization of the model within the NCG framework (in the sense of path integrals) is
yet to be fully explored,\cite{connes-book1,connes-book2} 
so one usually treats the NCG spectral action as an effective QFT action at the unification scale,
and evolve down to lower energies using the usual Renormalization Group (RG) equations to work out
the infrared consequences. 
\item The minimal version of NCG model which describes the SM predicts a Higgs mass of $\sim 170\;\mathrm{GeV}$,
in clear contradiction with experiment.\cite{Chamseddine:2010ud}. This issue could be remedied by turning one of the off-diagonal entries of the Dirac operator, which is responsible for the neutrino Majorana mass, into a field. With this singlet field coupled to the SM Higgs field, the model accommodates a $126\;\mathrm{GeV}$ Higgs boson.\cite{Chamseddine:2012sw}
This could also be accomplished by extending 
the NCG to that which leads to a left-right symmetric Pati-Salam type action
with coupling unification which automatically involves this singlet field.\cite{Chamseddine:2013rta}
\end{itemize}
In addition, the NCG spectral action approach to particle physics is under continued development by Connes and collaborators, and sorting out
the various versions can be difficult.

Despite these caveats, however, or any other reservation one may have about the entire approach,
it is not without its merits, as explained above, and we feel that it may have the potential to
develop into a full-fledged paradigm.
In particular, from the phenomenological standpoint,
the necessity to enlarge the gauge symmetry 
(via an enlargement of the underlying NCG)
to accommodate the Higg mass can be considered a strength rather than a weakness.
It tells us that the approach is restrictive enough for the models to be confronted by experiment,
and point us in new directions to explore.

Indeed, in a recent paper \cite{Chamseddine:2015ata}, 
Chamseddine, Connes, and van Suijlekom have proposed a new formulation of an NCG based
unified left-right symmetric Pati-Salam model, which comes in three different versions
differing in Higgs content. In all three, the gauge theory which emerges from the
underlying NCG at the unification scale, which we will call $M_U$, is that with gauge symmetry
$G_{224} = SU(2)_L\times SU(2)_R\times SU(4)_C$
with unified couplings:
\begin{equation}
g_L(M_U) \;=\; g_R(M_U) \;=\; g_4(M_U)\;.
\end{equation}
In one version, the symmetry is actually $G_{224D} = G_{224}\times D$, where $D$ denotes parity
which maintains left-right symmetry.\footnote{$D$-parity is slightly different from the usual Lorentz ($P$) parity in that the former interchanges the $SU(2)_L$ and $SU(2)_R$ sectors completely (including the scalars), while the latter does not transform the scalars. For example, the $D$-parity interchanges the $SU(2)_L$ Higgs fields and their $SU(2)_R$ counterparts, and transforms the bidoublet $\phi$ into $\phi^{\dagger}$ (and vice versa), while the $P$-parity leaves them unchanged.}

$G_{224}$ or $G_{224D}$ is assumed to break down to $G_{213}=SU(2)_L\times U(1)_Y\times SU(3)_C$
of the SM at scale $M_R$ with matching conditions
\begin{eqnarray}
\dfrac{1}{[g_1(M_R)]^2} & = & 
\dfrac{2}{3}\dfrac{1}{[g_4(M_R)]^2} +
\dfrac{1}{[g_R(M_R)]^2} \;,\cr
\dfrac{1}{[g_2(M_R)]^2} & = &
\dfrac{1}{[g_L(M_R)]^2} \;,\cr
\dfrac{1}{[g_3(M_R)]^2} & = & 
\dfrac{1}{[g_4(M_R)]^2} \;.
\label{MRMCmatching}
\end{eqnarray}
For all three versions, which differ in particle content, Ref.~\citen{Chamseddine:2015ata} argues that both
boundary conditions can be satisfied if $M_U\sim 10^{15}\,\mathrm{GeV}$ and $M_R\sim 10^{13}\,\mathrm{GeV}$.

In this paper, we will not attempt to review or justify the derivation of these models, but
look at their consequences purely phenomenologically.
From that viewpoint, the high value of $M_R$ is problematic in light
of recent hints of a $W_R$ with a mass of around $2\;\mathrm{TeV}$ at the LHC.\cite{Aad:2015owa,Khachatryan:2015sja,Aad:2014aqa,CMS:2015gla,Khachatryan:2014gha,ATLAS-CONF-2015-045}
If the LHC signal is indeed the gauge boson of the $SU(2)_R$ group, then $M_R$ on the order of a few TeV would be more compatible with that possibility.
For instance, in Refs.~\citen{Aydemir:2013zua,Aydemir:2014ama} we proposed an $su(2/2)$ superconnection-based left-right symmetric model
for which $M_R=4\,\mathrm{TeV}$, placing the mass of $W_R$ in the correct range.
We address the question whether $M_R$ for Chamseddine et al.'s NCG models can be lowered by 
the addition of intermediate breaking scales between $M_U$ and $M_R$ at which
the symmetry breaks down from $G_{224D}/G_{224}$ to $G_{213}$ via several intermediate steps.
In other words, is any symmetry breaking pattern compatible with a unified left-right symmetric
Pati-Salam model at $M_U$, and the SM below $M_R\sim\mbox{few TeV}$?
We will demand that $M_U$ stay below the Planck mass at $10^{19}\,\mathrm{GeV}$.
Similar analyses have been carried out in the context of non-supersymmetric
$SO(10)$ GUT models in Refs.~\citen{Chang:1984qr,Parida:1989an,Deshpande:1992au,Bertolini:2009qj,Babu:2012vc,Awasthi:2013ff,Bandyopadhyay:2015fka,Aydemir:2015oob}
for a variety of symmetry breaking chains.\footnote{%
For analyses of supersymmetric $SO(10)$ GUT, see Refs.~\citen{Deshpande:1992eu,Majee:2007uv,Parida:2008pu,Parida:2010wq}.
} 
Our analysis differs from these due to
the NCG models considered here differing in Higgs content 
since NCG does not require the Higgs fields to fall into $SO(10)$ multiplets.

While originally motivated by the desire to confront the viability of
NCG derived unified left-right symmetric Pati-Salam models, we note in passing that
similar models may emerge in a large class of string compactifications as discovered
by Dienes.\cite{Dienes:2006ut}
So the results presented here may have a wider range of applicability.

This paper is organized as follows.
In section~2, we briefly review the current status of the $W_R$ like signal seen at the LHC,
and what the phenomenological constraints are.
In section~3, we cover various symmetry breaking chains from $G_{224D}/G_{224}$ down to $G_{213}$,
and solve the renormalization group evolution equations for breaking scales which would satisfy
the boundary/matching conditions for the given particle content.
The list includes those that were considered by Chamseddine, Connes, and van Suijlekom in 
Ref.~\citen{Chamseddine:2015ata}.
We conclude in section~4 with a discussion of what was discovered.

\section{Status of the $W_R$ signal at the LHC}

Recently, ATLAS reported on a search for narrow resonances hadronically decaying into a pair of SM gauge bosons $WW$, $WZ$, or $ZZ$ \cite{Aad:2015owa}. The largest excess occurs in the $WZ$ channel at around $2$ TeV with a local significance of $3.4\sigma$ and a global $2.5\sigma$.  Moreover, both CMS \cite{Khachatryan:2015sja} and ATLAS \cite{Aad:2014aqa} notice an excess at around $1.8$ TeV in the dijet distributions albeit with low significance ($2.2\sigma$ and $1\sigma$). 
In addition, CMS observes an excess, again at around $2$ TeV, 
both in their search for massive $WH$ production in the $\ell\nu b \overline{b}$ final state \cite{CMS:2015gla} 
and in massive resonance production decaying into two SM vector bosons (one of which is leptonically tagged \cite{Khachatryan:2014gha}), 
both of which have lower significance than $2\sigma$. 

It is discussed in Refs.~\citen{Deppisch:2014qpa,Deppisch:2014zta,Brehmer:2015cia,Deppisch:2015cua} that these results may be interpreted in the context of the left-right model with the gauge group $SU(2)_L\times SU(2)_R\times U(1)'$ and it is shown that a heavy right-handed gauge boson  $W_R$ with a single coupling $g_R(M_R)\simeq 0.4$ can explain the current measurements. Note that this coupling is different from the SM left-handed $W_L$ coupling $g_L(5 \mbox{TeV})\simeq 0.63$.\cite{Agashe:2014kda,ALEPH:2005ab} 
Many other authors have also discussed possible phenomenological consequences of the $W_R$ interpretation,
e.g. Refs.~\citen{Dobrescu:2015qna,Dobrescu:2015yba,Cacciapaglia:2015nga,Fukano:2015uga,Chen:2015xql,Omura:2015nwa,Lane:2015fza,Arnan:2015csa,Dev:2015pga,Coloma:2015una,Aguilar-Saavedra:2015rna,Heikinheimo:2014tba} to list just a few,
but we refrain from reviewing them here.

\section{TeV-scale left-right model in the light of latest LHC searches}

\subsection{Setup of the Problem}

We would like to see whether such a $W_R$ can be accommodated within an NCG induced 
unified left-right symmetric Pati-Salam model.
The left-right symmetric model naturally has $g_R=g_L$. However, one can have an asymmetry between $g_R$ and $g_L$ if one separates the $D$-parity\cite{Maiezza:2010ic} breaking scale $M_D$ from the the scale $M_R$ where $SU(2)_R$ is broken \cite{Chang:1983fu,Chang:1984uy}. 

As an intermediate symmetry between $G_{224D}/G_{224}$ and $G_{213}$ of the SM, we introduce
\begin{equation}
G_{2213} \;=\; SU(2)_L\times SU(2)_R\times U(1)_{B-L} \times SU(3)_C\;,
\end{equation}
with gauge couplings $g_L$, $g_R$, $g_{BL}$, and $g_3$.
The most general breaking sequence will then be
\begin{equation}
\mathrm{NCG} 
\;\xRightarrow{M_U}\; G_{224D} 
\;\xrightarrow{M_D}\; G_{224} 
\;\xrightarrow{M_C}\; G_{2213}
\;\xrightarrow{M_R}\; G_{213} 
\;\xrightarrow{M_Z}\; G_{13}\;,
\label{generalNCGbreakingsequence}
\end{equation} 
where the double-line arrow indicates the emergence of the $G_{224D}$ theory from the underlying NCG,
and $G_{13}=U(1)_{\mathrm{EM}}\times SU(3)_C$ is the unbroken group which remains below the electroweak scale
with couplings $e$ and $g_3$.

We label the energy intervals in between symmetry breaking scales
starting from $[M_Z,M_R]$ up to $[M_D,M_U]$ with Roman numerals as: 
\begin{eqnarray}
\mathrm{I}   & \;:\; & [M_Z,\;M_R]\;,\quad G_{213}  \;(\mathrm{SM}) \;,\cr
\mathrm{II}  & \;:\; & [M_R,\;M_C]\;,\quad G_{2213} \;,\cr
\mathrm{III} & \;:\; & [M_C,\;M_D]\;,\quad G_{224}  \;,\cr
\mathrm{IV}  & \;:\; & [M_D,\;M_U]\;,\quad G_{224D} \;.
\label{IntervalNumber}
\end{eqnarray}
The ordering of the breaking scales must be strictly maintained, that is
\begin{equation}
M_Z \;\le\; M_R \;\le\; M_C \;\le\; M_D \;\le\; M_U\;.
\end{equation}
However, adjacent scales can be set equal which collapses the corresponding energy interval 
and skips the intermediate step in between.
For instance, if $M_R=M_C$, then $G_{224}$ breaks directly to $G_{213}$, and
interval III will be followed by interval I, skipping interval II.

In the following, we will investigate whether it is possible to set $M_R\sim 5\,\mathrm{TeV}$, 
while maintaining $M_U$ below the Planck scale.
The IR data which we will keep fixed as boundary conditions to the RG running are 
\cite{Agashe:2014kda,ALEPH:2005ab}
\begin{eqnarray}
\alpha(M_Z) & = & 1/127.9\;,\cr
\alpha_s(M_Z) & = & 0.118\;,\cr
\sin^2\theta_W(M_Z) & = & 
0.2312\;,
\label{SMboundary}
\end{eqnarray}
at $M_Z=91.1876\,\mathrm{GeV}$, which translates to
\begin{equation}
g_1(M_Z) \;=\; 0.36\;,\quad
g_2(M_Z) \;=\; 0.65\;,\quad
g_3(M_Z) \;=\; 1.22\;.
\label{MZboundary}
\end{equation}
The coupling constants are all required to remain in the perturbative regime during the
evolution from $M_U$ down to $M_Z$.

\begin{center}
\begin{table}[b]
\tbl{Dynkin index $T_i$ for several irreducible representations of $SU(2)$, $SU(3)$, and $SU(4)$. 
Different normalization conventions are used in the literature. 
For example, there is a factor of 2 difference between those given in Ref.~\citen{Lindner:1996tf} and those in Ref.~\citen{Slansky:1981yr}. Our convention follows the former.
For $SU(3)$, there exist two inequivalent 15 dimensional irreducible representations.
}
{\begin{tabular}{ccccc}
\toprule
\ \ \ Representation\ \ \ \ \ & $\qquad SU(2)\qquad$ & $\qquad SU(3)\qquad$ & $\qquad SU(4)\qquad$ \\ 
\colrule
$\vphantom{\bigg|}$ 2 &   $\dfrac{1}{2}$ &              $-$ &   $-$ & $\vphantom{\bigg|}$ \\
$\vphantom{\bigg|}$ 3 &                2 &   $\dfrac{1}{2}$ &   $-$ & $\vphantom{\bigg|}$ \\
$\vphantom{\bigg|}$ 4 &                5 &              $-$ &   $\dfrac{1}{2}$ & $\phantom{\bigg|}$ \\ 
$\vphantom{\bigg|}$ 6 &  $\dfrac{35}{2}$ &   $\dfrac{5}{2}$ &   $1$ & $\vphantom{\bigg|}$ \\
$\vphantom{\bigg|}$ 8 &               42 &              $3$ &   $-$ & $\vphantom{\bigg|}$ \\
$\vphantom{\bigg|}$10 & $\dfrac{165}{2}$ &  $\dfrac{15}{2}$ &   $3$ & $\vphantom{\bigg|}$ \\
$\vphantom{\bigg|}$15 &              280 & $10,\dfrac{35}{2}$ &  4 & $\vphantom{\bigg|}$ \\
\botrule
\end{tabular}
\label{DynkinIndex}}
\end{table}
\end{center}

\subsection{One-Loop Running and the Extended Survival Hypothesis}

For a given particle content, the gauge couplings are evolved according to the 1-loop
RG relation
\begin{equation}
\frac{1}{g_{i}^{2}(M_A)} - \dfrac{1}{g_{i}^2(M_B)}
\;=\; \dfrac{a_i}{8 \pi^2}\ln\dfrac{M_B}{M_A}
\;,
\end{equation}
where the RG coefficients $a_i$ are given by \cite{Jones:1981we,Lindner:1996tf}
\begin{eqnarray}
\label{1loopgeneral}
a_{i}
\;=\; -\frac{11}{3}C_{2}(G_i)
& + & \frac{2}{3}\sum_{R_f} T_i(R_f)\cdot d_1(R_f)\cdots d_n(R_f) \cr
& + & \frac{1}{3}\sum_{R_s} T_i(R_s)\cdot d_1(R_s)\cdots d_n(R_s)\;.
\end{eqnarray}
Here, the summation is over irreducible chiral representations of fermions ($R_f$) in the second term and those of scalars ($R_s$) in the third. 
$C_2(G_i)$ is the quadratic Casimir for the adjoint representation of the group $G_i$, 
and $T_i$ is the Dynkin index of each (complex) representation.\footnote{If the representation is real a factor of $\frac{1}{2}$ comes about in the third term.} 
For $SU(2)$, $C_2(G)=2$, $T=1/2$ for doublet representations and $T=2$ for triplets. 
See Table~\ref{DynkinIndex} for the Dynkin indexes of other representations.
For $U(1)$, $C_2(G)=0$ and
\begin{equation}
\sum_{f,s}T \;=\; \sum_{f,s}\left(\dfrac{Y}{2}\right)^2\;,
\label{U1Dynkin}
\end{equation}
where $Y/2$ is the $U(1)$ charge, the factor of $1/2$ coming from the traditional 
normalizations of the hypercharge $Y$ and $B-L$ charges.
The $a_i$'s will differ depending on the particle content, which changes every time symmetry breaking occurs.
We will distinguish the $a_i$'s in different energy intervals with the corresponding roman numeral superscript,
cf. Eq.~(\ref{IntervalNumber}).

For the particle content in each energy interval we impose the \textit{Extended Survival Hypothesis} (ESH). \cite{delAguila:1980at}
ESH states that at every step of the symmetry breaking chain, the only scalars which survive below the symmetry breaking scale are the ones which acquire vacuum expectation values (VEV's) at the subsequent levels of the symmetry breaking. 
For instance, the only scalar assumed to survive below $M_R$ would be the SM Higgs doublet which acquires a VEV to
break $G_{213}$ further down to $G_{13}$ under the ESH.

\subsection{Non-Unified Left-Right Symmetric Pati-Salam}

We begin by looking at the Pati-Salam model\cite{Pati:1974yy, Senjanovic:1975rk, Mohapatra:1974gc, Mohapatra:1974hk} without the unification of all three couplings as demanded in the
NCG approach.  We impose left-right symmetry $g_L=g_R$ at scale $M_D$,
which we identify as the scale at which $G_{224D}$ breaks to $G_{224}$, and
evolve our couplings down from $M_D$:
\begin{equation}
G_{224D} 
\;\xrightarrow{M_D}\; G_{224} 
\;\xrightarrow{M_C}\; G_{2213}
\;\xrightarrow{M_R}\; G_{213} 
\;\xrightarrow{M_Z}\; G_{13}\;.
\end{equation} 
Note that energy interval IV is absent.
In addition to Eq.~(\ref{MZboundary}),
the boundary/matching conditions we impose on the couplings at the symmetry breaking scales are:
\begin{eqnarray}
M_D & \;:\; &  g_L(M_D) \;=\; g_R(M_D) \;, \vphantom{\Big|} \label{MPmathcing}\\
M_C & \;:\; & \sqrt{\frac{2}{3}}\,g_{BL}(M_C) \;=\; g_3(M_C)=g_4(M_C) \;,\label{MCmatching}\\
M_R & \;:\; & \frac{1}{g_1^2(M_R)} \;=\; \frac{1}{g_R^2(M_R)}+\frac{1}{g_{BL}^2(M_R)}\;,\quad
g_2(M_R)\;=\;g_L(M_R)\;, \label{MRmatching} \\
M_Z & \;:\; & \frac{1}{e^2(M_Z)} \;=\; \frac{1}{g_1^2(M_Z)}+\frac{1}{g_2^2(M_Z)}\;.
\label{MZmatching}
\end{eqnarray}
Note that if $M_C=M_R$, then the conditions at those scales reduce to those given
in Eq.~(\ref{MRMCmatching}).

We assume that the above breaking sequence is accomplished by a Higgs sector consisting of scalars
which transform under $G_{224}$ as
\begin{equation}
\phi(2,2,1)\;,\quad\Delta_R (1,3,10)\;,\quad\Sigma(1,1,15)\;.
\end{equation}
These fields decompose into irreducible representations of $G_{2213}$ as:
\begin{eqnarray}
\Sigma(1,1,15) 
& = & \Sigma_1(1,1,0,1) 
\oplus \Sigma_3\left(1,1,\dfrac{4}{3},3\right) 
\oplus \Sigma_{\bar{3}}\left(1,1,\dfrac{-4}{3},\bar{3}\right) 
\oplus \Sigma_8(1,1,0,8)
\;,\cr
\Delta_R (1,3,10)
& = & \Delta_{R1}(1,3,2,1) 
\oplus \Delta_{R3}\left(1,3,\frac{2}{3},3\right) 
\oplus \Delta_{R6}\left(1,3,\frac{-2}{3},6\right)
\;, \cr
\phi(2,2,1) & = & \phi(2,2,0,1)\;.\vphantom{\bigg|}
\label{SigmaDeltaphiDecomposition}
\end{eqnarray}
The breaking of $G_{224}$ down to $G_{2213}$ would be
accomplished by the field $\Sigma_1$ acquiring a VEV.
$\Sigma_3$, $\Sigma_{\bar{3}}$, $\Sigma_8$, $\Delta_{R3}$, $\Delta_{R6}$ are all colored, so they will not be acquiring VEV's
in the subsequent steps.
Thus, under the ESH, all these fields will become heavy at $M_C$ and decouple from the RG equations
below $M_C$.
The remaining fields decompose into irreducible representations of $G_{213}$ as:
\begin{eqnarray}
\Delta_{R1}(1,3,2,1) & = & \Delta_{R1}^{0}(1,0,1) \oplus \Delta_{R1}^{+}(1,2,1) \oplus \Delta_{R1}^{++}(1,4,1) \;,\cr
\phi(2,2,0,1) & = & \phi_2(2,1,1) \oplus \phi'_2 (2,-1,1) \;. \vphantom{\bigg|}
\label{Sigma1phiDecomposition}
\end{eqnarray}
The breaking of $G_{2213}$ down to $G_{213}$ would be accomplished by the field $\Delta_{R1}^{0}$, while that of $G_{213}$ down to $G_{13}$ would be realised by the neutral (diagonal) components of $\phi_2(2,2,0,1)$, acquiring VEVs. The fields $\Delta_{R1}^{+}$ and $\Delta_{R1}^{++}$ would be both charged under electromagnetism, so they will not be acquiring VEV's in the subsequent steps. Thus, under the ESH, these fields will become heavy at $M_R$. In addition, only one of the two physical states (which are linear combinations of  $\phi_2$ and $\phi'_2$) remains light while the other picks a mass at $M_R$, unless we apply fine-tuning \cite{Mohapatra:1982aq}. The left-over field, the SM Higgs (which can be identified without loss of generality as $ \phi_2(2,1,1)$)  is left to be the only field in the Higgs spectrum below $M_R$.
Thus, under the ESH, the particle content (other than the fermions and gauge bosons) of our model 
in the three energy intervals I through III are:
\begin{eqnarray}
\mathrm{III} & \;:\; & \phi(2,2,1)\;,\;\Delta_R (1,3,10)\;,\;\Sigma(1,1,15)\;,\cr
\mathrm{II}  & \;:\; & \phi(2,2,0,1)\;,\;\Delta_{R1}(1,3,2,1)\;, \vphantom{\bigg|}\cr
\mathrm{I}   & \;:\; & \phi_2(2,1,1)\;.
\end{eqnarray}
The values of the RG coefficients for this Higgs content are listed in Table~\ref{PSa1}.

\begin{table}[t]
\tbl{The Higgs content and RG coefficients in the three energy intervals for the 
non-unified left-right symmetric Pati-Salam model
under the Extended Survival Hypothesis (ESH).}
{\begin{tabular}{c|l|l}
\hline
$\vphantom{\Big|}$ Interval & Higgs content & RG coefficients 
\\
\hline
$\vphantom{\Biggl|}$ III   
& $\phi(2,2,1),\;\Delta_R (1,3,10),\;\Sigma(1,1,15)$ 
& $\left( a_{L},a_{R},a_{4}\right)^\mathrm{III}
 =\left(-3,\dfrac{11}{3},-7\right)$ 
\\
\hline
$\vphantom{\Biggl|}$  II  
& $\phi(2,2,0,1),\;\Delta_{R1}(1,3,2,1)$ 
& $\left(a_{L},a_{R},a_{BL},a_3\right)^\mathrm{II}=\left(-3,\dfrac{-7}{3},\dfrac{11}{3},-7\right)$
\\
\hline
$\vphantom{\Biggl|}$   I  
& $\phi_2(2,1,1)$ 
& $\left(a_{1},a_{2},a_{3}\right)^\mathrm{I}=\left(\dfrac{41}{6},\dfrac{-19}{6},-7\right)$ 
\\
\hline
\end{tabular}}
\label{PSa1}
\end{table}

Taking advantage of the boundary/matching conditions,
the following relations can be derived between the boundary values $\alpha(M_Z)$,
$\alpha_s(M_Z)$, $\sin^2\theta_W(M_Z)$, $g_R(M_R)$, and the ratios of the successive symmetry
breaking scales:
\begin{eqnarray}
\lefteqn{
2\pi
\left[
\dfrac{3-6\sin^2\theta_W(M_Z)}{\alpha(M_Z)}-\dfrac{2}{\alpha_s(M_Z)}
\right]
}
\cr
& = & 
 \underbrace{\left(3a_1-3a_2-2a_3\right)^\mathrm{I}}_{44}\;\ln\dfrac{M_R}{M_Z}
+\underbrace{\left(-3a_L+3a_R+3a_{BL}-2a_3\right)^\mathrm{II}}_{27}\;\ln\dfrac{M_C}{M_R}
\cr & & \quad
+\underbrace{\left(-3a_L+3a_R\right)^\mathrm{III}}_{20}\;\ln\dfrac{M_D}{M_C}
\;,
\cr
\lefteqn{
2\pi\left[\dfrac{4\pi}{g_R^2(M_R)} - \dfrac{\sin^2\theta_W(M_Z)}{\alpha(M_Z)}\right]
} \cr
& = & 
 \underbrace{(-a_2)^\mathrm{I}}_{19/6}\;\ln\dfrac{M_R}{M_Z}
+\underbrace{\left(a_R - a_L\right)^\mathrm{II}}_{2/3}\;\ln\dfrac{M_C}{M_R}
+\underbrace{\left(a_R - a_L\right)^\mathrm{III}}_{20/3}\;\ln\dfrac{M_D}{M_C}
\;.
\end{eqnarray}
The derivation is shown in the Appendix.
To maintain the ordering of the mass scales, all logarithms in these
expressions must be non-negative.  
Numerically, we have
\begin{eqnarray}
517 & = & 44\,x + 27\,y + 20\,z \;,\cr
\dfrac{206}{g_R^2(M_R)}-484
& = & 19\,x + \hphantom{0}4\,y + 40\,z\;,
\end{eqnarray}
where
\begin{equation}
x\;=\;\log_{10}\dfrac{M_R}{M_Z}\;,\qquad
y\;=\;\log_{10}\dfrac{M_C}{M_R}\;,\qquad
z\;=\;\log_{10}\dfrac{M_D}{M_C}\;.
\label{xyzDef}
\end{equation}
If we fix $M_R=5\,\mathrm{TeV}$, then
$x=\log_{10}(M_R/M_Z)=1.74$, and the above system of linear
equations yields
\begin{equation}
y\;=\;27.9-\dfrac{4.11}{g_R^2(M_R)}\;,\qquad
z\;=\;-15.7+\dfrac{5.56}{g_R^2(M_R)}
\;.
\end{equation}
Since both $y$ and $z$ must be positive, 
we must have
\begin{equation}
0.38\;<\; g_R(M_R) \;<\; 0.59\;.
\label{gRconstraint1}
\end{equation}
We would also like to impose the condition
\begin{equation}
x+y+z\;=\;14.0+\dfrac{1.44}{g_R^2(M_R)}\;=\;\log_{10}\dfrac{M_D}{M_Z} \;<\;
\log_{10}\dfrac{10^{19}\,\mathrm{GeV}}{M_Z}
\;=\; 17.0\;,
\end{equation}
which constrains $g_R(M_R)$ to
\begin{equation}
g_R(M_R)\;>\;0.69\;,
\label{gRconstraint2}
\end{equation}
which is incompatible with Eq.~(\ref{gRconstraint1}).
Thus, the system does not allow for a parity breaking scale $M_D$ lower than the
Planck mass.  

\begin{figure}[t]
\subfigure[$g_R(M_R)=0.4$]{\includegraphics[width=6.5cm]{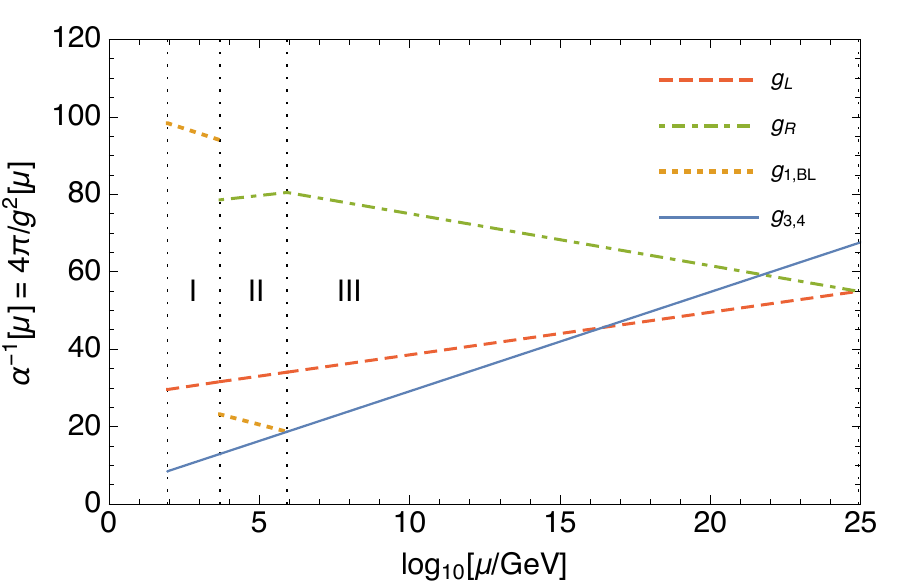}}\subfigure[$g_R(M_R)=0.59$]{\includegraphics[width=6.5cm]{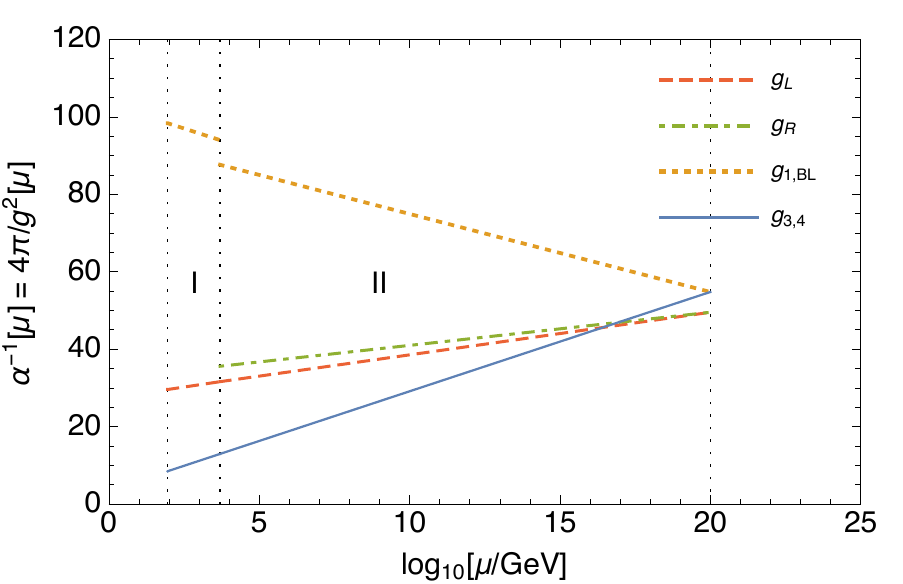}}
\caption{Running of the gauge couplings for the left-right symmetric
Pati-Salam model.  The vertical dotted lines from left to right correspond to the
symmetry breaking scales $M_Z$, $M_R$, and $M_C$.
$M_R$ is fixed at 5 TeV.
For the $U(1)_{B-L}$ coupling between $M_R$ and $M_C$, we plot
$\dfrac{3}{2}\alpha^{-1}_{BL}(\mu)=\dfrac{6\pi}{g_{BL}^2(\mu)}$  
so that it agrees with $\alpha_4^{-1}(\mu)$ at $\mu=M_C$.
The two cases shown are (a) $g_R(M_R)=0.4$ is imposed, and
(b) $M_D$ is minimized by collapsing the energy interval III.
}
\label{RGrunning1}
\end{figure}

Indeed, if we set $g_R(M_R)=0.4$ as preferred by experiment \cite{Deppisch:2014qpa,Deppisch:2014zta,Brehmer:2015cia,Deppisch:2015cua},
we obtain $y=2.2$, $z=19.0$, which translates to
\begin{equation}
M_R\;=\; 5\,\mathrm{TeV}\;,\quad
M_C\;=\; 8\times 10^5\,\mathrm{GeV}\;,\quad
M_D\;=\; 8\times 10^{24}\,\mathrm{GeV}\;,
\end{equation}
with
\begin{equation}
g_L(M_D)\;=\;g_R(M_D)\;=\; 0.48 \;,\qquad
g_4(M_D)\;=\; 0.43\;.
\end{equation}
If we allow $g_R(M_R)$ to be as large as $0.59$, we obtain
$y=16.3$, $z=0$, which translates to
\begin{equation}
M_R\;=\; 5\,\mathrm{TeV}\;,\quad
M_C\;=\; M_D\;=\; 1\times 10^{20}\,\mathrm{GeV}\;,
\end{equation}
with
\begin{equation}
g_L(M_D)\;=\;g_R(M_D)\;=\; 0.50 \;,\qquad
g_4(M_D)\;=\; 0.48\;.
\end{equation}
The evolution of the couplings for these choices of scales are shown in Fig.~\ref{RGrunning1}.
For each choice of $g_R(M_R)$, the value of $g_{BL}(M_R)$ is determined from the known value of
the hypercharge coupling $g_1(M_R)$ and the matching condition Eq.~(\ref{MRmatching}).
Larger values of $g_R(M_R)$ closer to $g_L(M_R)$ will lower the scale $M_D$ at which the
RG flow of the two couplings separate. 
At the same time, larger values of $g_R(M_R)$ demand smaller values of $g_{BL}(M_R)$, 
which pushes up the scale $M_C$ where the RG flow of $g_{BL}$ bifurcates from that of $g_3$.
Since the order $M_C\le M_D$ cannot be violated, $M_D$ cannot be lowered further by increasing $g_R(M_R)$
once the two scales meet.

Looking at Fig.~\ref{RGrunning1}(b), however, we notice that 
in energy interval II $g_L$ and $g_R$ do flow apart, but not as
much as in energy interval III.  A larger difference between $g_L$ and $g_R$ could
be generated in interval II if $(a_L-a_R)^\mathrm{II}$ could be enhanced.
To this end, let us relax the ESH and allow some of the colored $\Delta_R$ fields to
survive into interval II.
The RG coefficients for three Higgs-content scenarios in interval II
different from the ESH case are listed in Table~\ref{PSa2}.
Clearly, the addition of extra $\Delta_R$ fields enhances $(a_L-a_R)^\mathrm{II}$.

We perform the same analyses as above for the three ESH-breaking cases, namely, the calculation of the
symmetry breaking scales to reproduce $g_R(M_R)=0.4$, and then by allowing the value of
$g_R(M_R)$ to float in order to find the lowest value of $M_D$:

\begin{table}[t]
\tbl{The dependence of the RG coefficients on the Higgs content in energy interval II
where the symmetry is $G_{2213}$.  Relaxing the ESH will lead to different Higgs content
and different RG coefficients.
}
{\begin{tabular}{c|l|l}
\hline
$\vphantom{\bigg|}$ Interval & Higgs content & $\left(a_{L},a_{R},a_{BL},a_3\right)^\mathrm{II}$ \\
\hline
$\vphantom{\Biggl|}$  II 
& $\phi(2,2,0,1),\;\Delta_{R1}(1,3,2,1)$ 
& $\left(-3,\dfrac{-7}{3},\dfrac{11}{3},-7\right)$
\\
\cline{2-3}
& $\phi(2,2,0,1),\;\Delta_{R1}(1,3,2,1),\;\Delta_{R3}\left(1,3,\dfrac{2}{3},3\right)$
& $\left(-3,\dfrac{-1}{3},4,\dfrac{-13}{2}\right)$ $\vphantom{\Bigg|}$
\\
\cline{2-3}
& $\phi(2,2,0,1),\;\Delta_{R1}(1,3,2,1),\;\Delta_{R6}\left(1,3,\dfrac{-2}{3},6\right)$
& $\left(-3,\dfrac{5}{3},\dfrac{13}{3},\dfrac{-9}{2}\right)$ $\vphantom{\Bigg|}$
\\
\cline{2-3}
& $\phi(2,2,0,1),\;\Delta_{R1}(1,3,2,1),\;\Delta_{R3}\left(1,3,\dfrac{2}{3},3\right)\!,\;\Delta_{R6}\left(1,3,\dfrac{-2}{3},6\right)$
& $\left(-3,\dfrac{11}{3},\dfrac{14}{3},-4\right)$ $\vphantom{\Bigg|}$
\\
\hline
\end{tabular}}
\label{PSa2}
\end{table}

\begin{figure}[t]
\subfigure[$g_R(M_R)=0.4$]{\includegraphics[width=6.5cm]{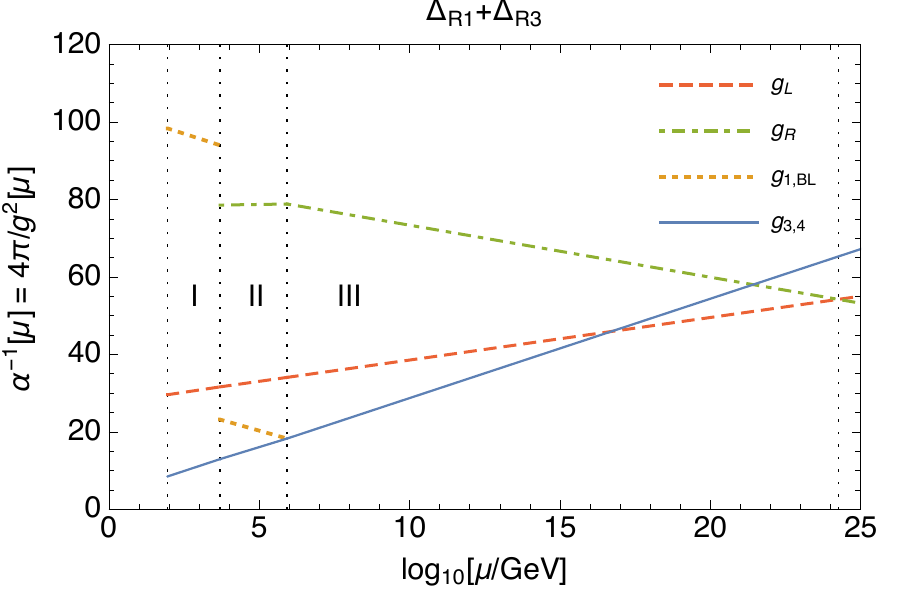}}\subfigure[$g_R(M_R)=0.53$]{\includegraphics[width=6.5cm]{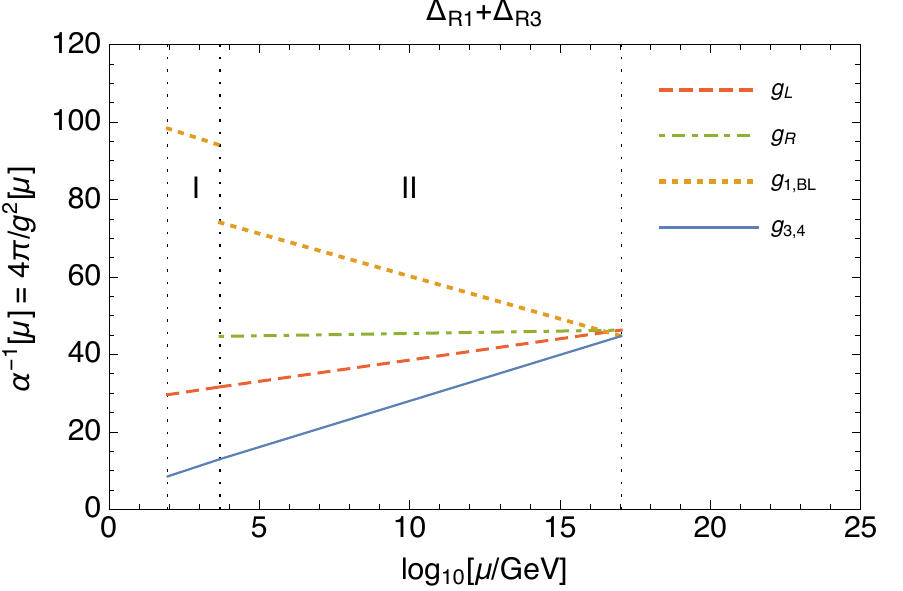}}
\subfigure[$g_R(M_R)=0.4$]{\includegraphics[width=6.5cm]{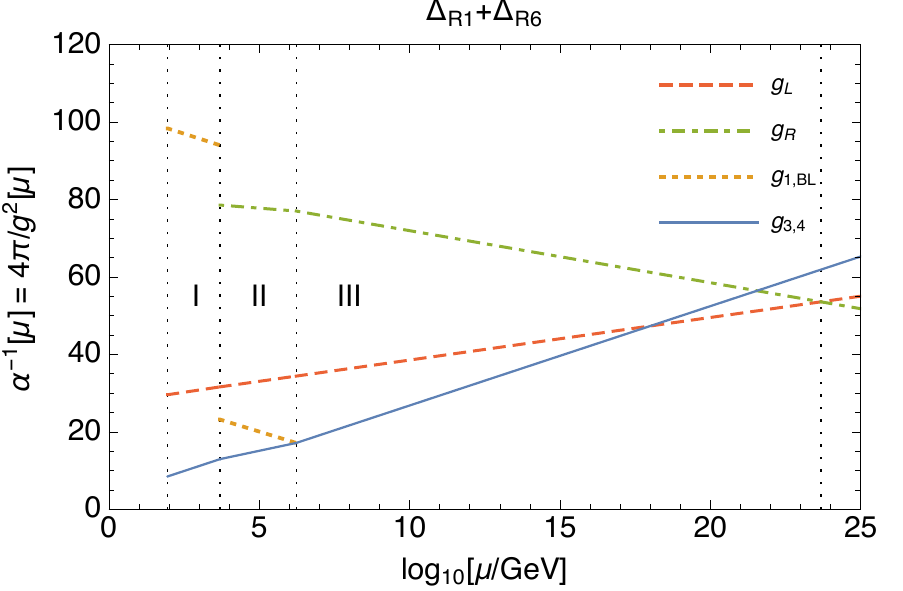}}\subfigure[$g_R(M_R)=0.49$]{\includegraphics[width=6.5cm]{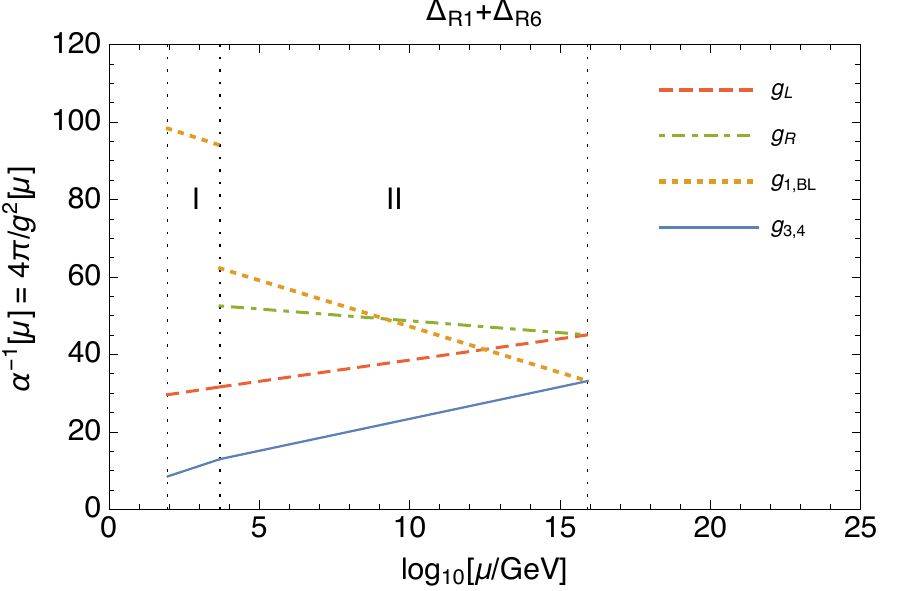}}
\subfigure[$g_R(M_R)=0.4$]{\includegraphics[width=6.5cm]{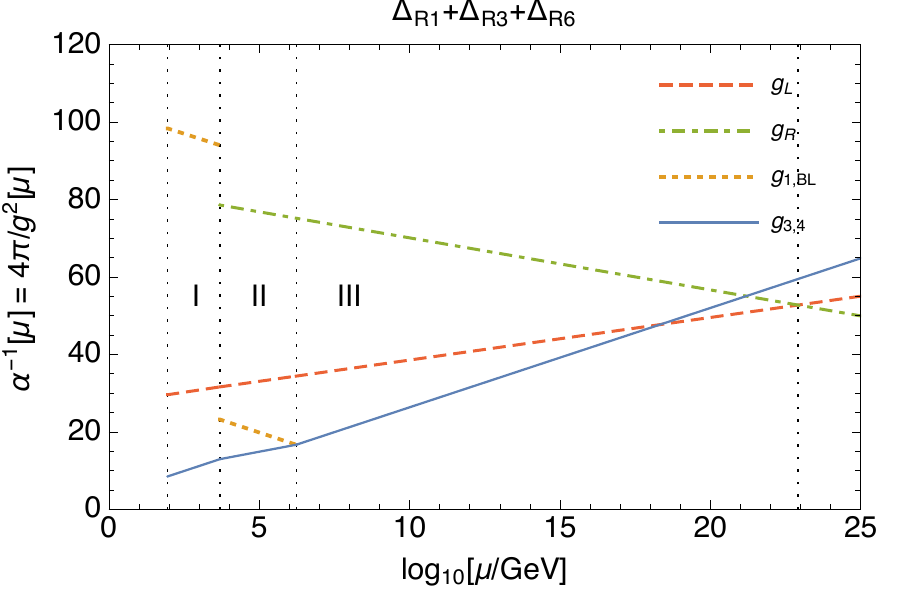}}\subfigure[$g_R(M_R)=0.47$]{\includegraphics[width=6.5cm]{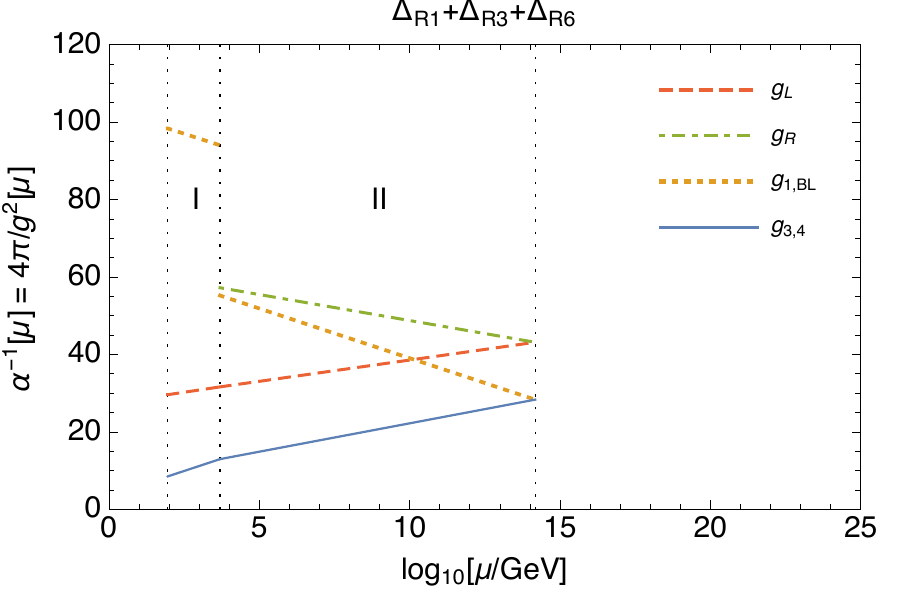}}
\caption{Running of the gauge couplings for the left-right symmetric
Pati-Salam model with more than $\Delta_{R1}$ surviving into 
energy interval II.
Vertical dotted lines indicate symmetry breaking scales.
$M_R$ is fixed at 5 TeV.
In (a), (c), and (e) $g_R(M_R)=0.4$ is imposed, while in
(b), (d), and (d) $M_D$ is minimized by collapsing the energy interval III.
}
\label{RGrunning2}
\end{figure}

\begin{enumerate}
\item $\Delta_{R1}$ and $\Delta_{R3}$ survive:\\
To reproduce $g_R(M_R)=0.4$, we find
\begin{equation}
M_R\;=\;5\,\mathrm{TeV}\;,\quad
M_C\;=\;8\times 10^{5}\,\mathrm{GeV}\;,\quad
M_D\;=\;2\times 10^{24}\,\mathrm{GeV}\;,
\end{equation}
with 
\begin{equation}
g_L(M_D)\;=\;g_R(M_D)\;=\; 0.48 \;,\qquad
g_4(M_D)\;=\; 0.44\;.
\end{equation}
If $g_R(M_R)$ is allowed to float, the minimum of $M_D$ is achieved when
$g_R(M_R)=0.53$ with
\begin{equation}
M_R\;=\;5\,\mathrm{TeV}\;,\quad
M_C\;=\;M_D\;=\;1\times 10^{17}\,\mathrm{GeV}\;,
\end{equation}
with 
\begin{equation}
g_L(M_D)\;=\;g_R(M_D)\;=\; 0.52 \;,\qquad
g_4(M_D)\;=\; 0.53\;.
\end{equation}
The runnings of the couplings for these cases are shown in Fig.~\ref{RGrunning2}(a) and (b).

\bigskip
\item  $\Delta_{R1}$ and $\Delta_{R6}$ survive:\\
To reproduce $g_R(M_R)=0.4$, we find
\begin{equation}
M_R\;=\;5\,\mathrm{TeV}\;,\quad
M_C\;=\;2\times 10^{6}\,\mathrm{GeV}\;,\quad
M_D\;=\;5\times 10^{23}\,\mathrm{GeV}\;,
\end{equation}
with 
\begin{equation}
g_L(M_D)\;=\;g_R(M_D)\;=\; 0.48 \;,\qquad
g_4(M_D)\;=\; 0.45\;.
\end{equation}
If $g_R(M_R)$ is allowed to float, the minimum of $M_D$ is achieved when
$g_R(M_R)=0.49$ with
\begin{equation}
M_R\;=\;5\,\mathrm{TeV}\;,\quad
M_C\;=\;M_D\;=\;8\times 10^{15}\,\mathrm{GeV}\;,
\end{equation}
with 
\begin{equation}
g_L(M_D)\;=\;g_R(M_D)\;=\; 0.53 \;,\qquad
g_4(M_D)\;=\; 0.62\;.
\end{equation}
The runnings of the couplings for these cases are shown in Fig.~\ref{RGrunning2}(c) and (d).

\bigskip
\item All three multiplets $\Delta_{R1}$, $\Delta_{R3}$, and $\Delta_{R6}$ survive:\\
To reproduce $g_R(M_R)=0.4$, we find
\begin{equation}
M_R\;=\;5\,\mathrm{TeV}\;,\quad
M_C\;=\;2\times 10^{6}\,\mathrm{GeV}\;,\quad
M_D\;=\;8\times 10^{22}\,\mathrm{GeV}\;,
\end{equation}
with 
\begin{equation}
g_L(M_D)\;=\;g_R(M_D)\;=\; 0.49 \;,\qquad
g_4(M_D)\;=\; 0.46\;.
\end{equation}
If $g_R(M_R)$ is allowed to float, the minimum of $M_D$ is achieved when
$g_R(M_R)=0.47$ with
\begin{equation}
M_R\;=\;5\,\mathrm{TeV}\;,\quad
M_C\;=\;M_D\;=\;2\times 10^{14}\,\mathrm{GeV}\;,
\end{equation}
with 
\begin{equation}
g_L(M_D)\;=\;g_R(M_D)\;=\; 0.54 \;,\qquad
g_4(M_D)\;=\; 0.67\;.
\end{equation}
The runnings of the couplings for these cases are shown in Fig.~\ref{RGrunning2}(e) and (f).

\end{enumerate}

These results indicate that achieving a value of $g_R(M_R)=0.4$ 
at $M_R=5\,\mathrm{TeV}$ is not trivial
in this model, requiring a very high value of the parity breaking scale $M_D$.
Lowering this scale below the Planck mass cannot be achieved with the minimal
Higgs content in energy interval II as required by the ESH even if the value of $g_R(M_R)$
were allowed to float.
If one relaxes the ESH, then $M_D$ lower than the Planck mass is possible
provided $g_R(M_R)$ is allowed to be as large as $\sim 0.5$.
It is also preferable for $M_C$ and $M_D$ to be degenerate,
that is, for $G_{224D}$ to break directly to $G_{2213}$.

\subsection{Unified Left-Right Symmetric Pati-Salam from NCG}

With the above results in mind,
let us now look at the unified left-right symmetric Pati-Salam model which we expect to
emerge from an underlying NCG.  The breaking pattern now includes an emergence/unification scale
as in Eq.~(\ref{generalNCGbreakingsequence}),
and all four energy intervals listed in Eq.~(\ref{IntervalNumber}) must be taken into account
with an extra boundary condition at $M_U$:
\begin{eqnarray}
M_U & \;:\; & g_L(M_U) \;=\; g_R(M_U) \;=\; g_4(M_U) \;.
\label{MUmatching}
\end{eqnarray}
This leads to the relations
\begin{eqnarray}
\lefteqn{
2\pi\left[\dfrac{3-8\sin^2\theta_W(M_Z)}{\alpha(M_Z)}\right]
}\cr
& = & 
 \left(3a_1 -5a_2\right)^\mathrm{I}\,\ln\dfrac{M_R}{M_Z}
+\left(-5a_L + 3a_R + 3a_{BL}\right)^\mathrm{II}\,\ln\dfrac{M_C}{M_R}
\cr
& &
+\left(-5a_L+3a_R+2a_4\right)^\mathrm{III}\,\ln\dfrac{M_D}{M_C}
+\left(-5a_L+3a_R+2a_4\right)^\mathrm{IV}\,\ln\dfrac{M_U}{M_D}
\;,
\label{NCGeq1} \\
\lefteqn{
2\pi\left[\dfrac{3}{\alpha(M_Z)} - \dfrac{8}{\alpha_s(M_Z)}\right]
}\cr
& = &
 \left(3a_1 + 3a_2 - 8a_3\right)^\mathrm{I}\,\ln\dfrac{M_R}{M_Z}
+\left(3a_L + 3a_R + 3a_{BL} - 8a_3\right)^\mathrm{II}\,\ln\dfrac{M_C}{M_R}
\cr
& & 
+\left(3a_L+3a_R-6a_4\right)^\mathrm{III}\,\ln\dfrac{M_D}{M_C}
+\left(3a_L+3a_R-6a_4\right)^\mathrm{IV}\,\ln\dfrac{M_U}{M_D}
\;,
\label{NCGeq2} \\
\lefteqn{
2\pi\left[\dfrac{4\pi}{g_R^2(M_R)}-\dfrac{\sin^2\theta_W(M_Z)}{\alpha(M_Z)}\right]
} \cr
\cr
& = & 
 (-a_2^\mathrm{I})\,\ln\dfrac{M_R}{M_Z}
+\left(a_R - a_L\right)^\mathrm{II}\,\ln\dfrac{M_C}{M_R}
+\left(a_R - a_L\right)^\mathrm{III}\,\ln\dfrac{M_D}{M_C}
\;.
\label{NCGeq3}
\end{eqnarray}
The derivation is given in the Appendix.
Note that there is no $\ln M_U/M_D$ term in the last line
since parity is not broken in energy interval IV and $a_L^\mathrm{IV}=a_R^\mathrm{IV}$.
We will now look at the three models of Chamseddine, Connes, and van Suijlekom
in Ref.~\citen{Chamseddine:2015ata} one by one.

\renewcommand{\theenumi}{\Alph{enumi}}
\begin{enumerate}

\bigskip
\item{Pati-Salam with ``composite" Higgs fields}\footnote{%
The terminology of Ref.~\citen{Chamseddine:2015ata} may cause confusion with the standard concept of compositeness that is found in the literature. What the authors of Ref.~\citen{Chamseddine:2015ata} call a ``composite'' field seems to be
a field which does not transform under a single irreducible representation.}
\medskip

\begin{table}[t]
\tbl{Higgs content of Model A of Ref.~\citen{Chamseddine:2015ata}.
In Ref.~\citen{Chamseddine:2015ata}, the model emerges with symmetry $G_{224}$ at
$M_U=M_D$.  This breaks directly to $G_{213}$ of the SM at $M_C=M_R$.
We modify this process by allowing $M_C\neq M_R$, inserting energy interval II
with symmetry $G_{2213}$ between intervals III and I.
The Higgs content in interval II is based on the ESH.}
{\begin{tabular}{c|l|l}
\hline
$\vphantom{\Big|}$ Interval & Higgs content & RG coefficients \\
\hline
$\vphantom{\Biggl|}$ III 
& $\phi(2,2,1),\;\Delta_R (1,2,4),\;\Sigma(1,1,15)$ 
& $\left( a_{L},a_{R},a_{4}\right)^\mathrm{III}
 =\left(-3,\dfrac{-7}{3},\dfrac{-29}{3}\right)$ 
\\
\hline
$\vphantom{\Biggl|}$ II  
& $\phi(2,2,0,1),\;\Delta_{R1}(1,2,1,1)$ 
& $\left(a_{L},a_{R},a_{BL},a_3\right)^\mathrm{II}=\left(-3,\dfrac{-17}{6},\dfrac{17}{6},-7\right)$
\\
\hline
$\vphantom{\Biggl|}$ I 
& $\phi_2(2,1,1)$ 
& $\left(a_{1},a_{2},a_{3}\right)^\mathrm{I}=\left(\dfrac{41}{6},\dfrac{-19}{6},-7\right)$ 
\\
\hline
\end{tabular}}
\label{NCG-Model-A}
\end{table}

The first model of Ref.~\citen{Chamseddine:2015ata} emerges with symmetry $G_{224}$
at $M_U=M_D$, which breaks directly to $G_{213}$ of the SM at $M_C=M_R$.
So only energy intervals I and III are present.
The Higgs content of this model in interval III, as specified in
Ref.~\citen{Chamseddine:2015ata}, is in shown in Table~\ref{NCG-Model-A}.
We make a slight modification by taking the $\Sigma(1,1,15)$ field to be real, 
conforming to standard Pati-Salam literature, whereas
Ref.~\citen{Chamseddine:2015ata} assumes it to be complex.

In this case, 
Eqs.~(\ref{NCGeq1}) through (\ref{NCGeq3}) simply reduce to
those with the II and IV terms missing.
Then, the system of three equations has three unknowns, namely $\ln\dfrac{M_D}{M_C}$, $\ln\dfrac{M_R}{M_Z}$,
and $g_R(M_R)$, which allows us to determine all three.
We find:
\begin{eqnarray}
g_R(M_R) & = & 0.54\;,\cr
M_R\;=\;M_C & = & 4.1\times 10^{13}\,\mathrm{GeV}\;,\cr
M_U\;=\;M_D & = & 3.5\times 10^{15}\,\mathrm{GeV}\;,
\end{eqnarray}
in agreement with Ref.~\citen{Chamseddine:2015ata}.
The unified coupling in this case is
$g_L(M_U)=g_R(M_U)=g_4(M_U)=0.53$.
The running of the couplings for this case is shown in Fig.~\ref{RGrunning3}(a).

\bigskip

We now allow for $M_C\neq M_R$ and insert the energy interval II with symmetry $G_{2213}$.
To determine the Higgs content in this interval, we again invoke the ESH.
The decomposition of $\Sigma(1,1,15)$ into irreducible representations of $G_{2213}$ 
was given in Eq.~(\ref{SigmaDeltaphiDecomposition}) and it was concluded that all the components
of $\Sigma(1,1,15)$ become heavy and decouple from the RG equations at $M_C$.
The decomposition of $\Delta_R(1,2,4)$ into irreducible representations of $G_{2213}$ 
is given by
\begin{equation}
\Delta_R(1,2,4) \;=\;
\Delta_{R1}(1,2,1,1) \oplus \Delta_{R3}\left(1,2,\dfrac{-1}{3},3\right)\;.
\end{equation}
$\Delta_{R3}$ is colored so again by ESH it will become heavy and only $\Delta_{R1}$ will survive
into II.
The decomposition of $\Delta_{R1}$ into irreducible representations of $G_{213}$ is given by
\begin{equation}
\Delta_{R1}(1,2,1,1) \;=\;
\Delta_{R1}^0(1,0,1) \oplus \Delta_{R1}^+(1,1,1)
\;.
\end{equation}
The breaking of $G_{2213}$ down to $G_{213}$ would be accomplished by the field $\Delta_{R1}^{0}$ acquiring a VEV,
while $\Delta_{R1}^+$ has electromagnetic charge so it must become heavy.
The survival of $\phi_2(2,1,1)$ into I is as before.
Thus, the Higgs content of the model is as shown in Table~\ref{NCG-Model-A}.

\begin{figure}[t]
\subfigure[$g_R(M_R)=0.54$]{\includegraphics[width=6.5cm]{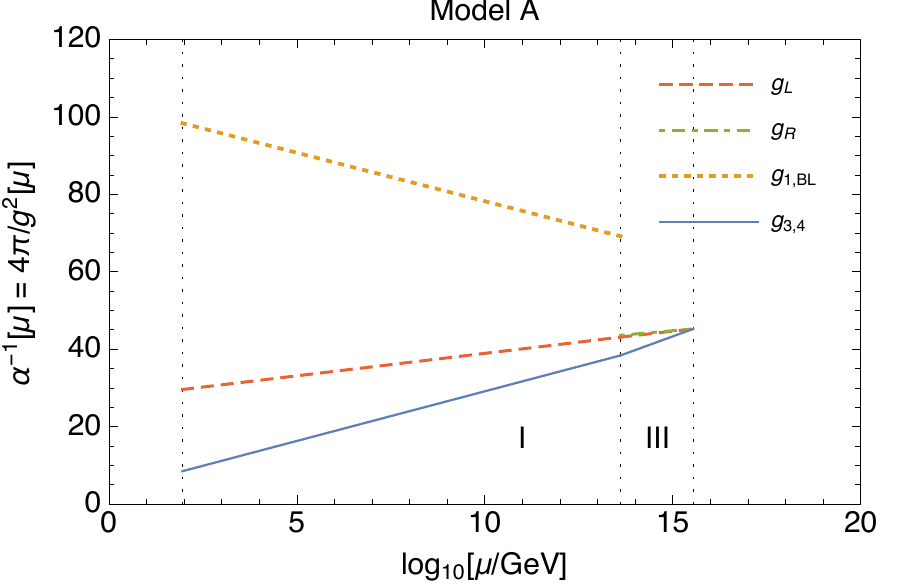}}\subfigure[$g_R(M_R)=0.58$]{\includegraphics[width=6.5cm]{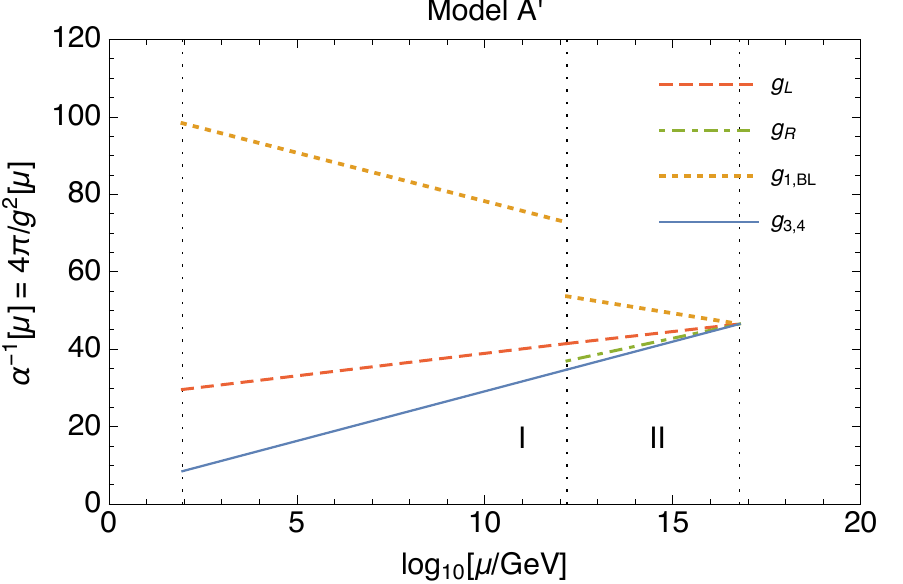}}
\caption{Running of the gauge couplings for Model A of Ref.~\citen{Chamseddine:2015ata}.
with (a) $G_{224}$ breaking directly into $G_{213}$, and 
(b) $G_{224}$ breaks immediately to $G_{2213}$ as it emerges.
In (a), the dashed line indicating $g_L$ and the dot-dashed line indicating $g_R$
are almost overlapping in interval III.
}
\label{RGrunning3}
\end{figure}

Eqs.~(\ref{NCGeq1}) through (\ref{NCGeq3}) now has four unknowns instead of three.
Numerically, they are given by
\begin{eqnarray}
401 & = & \dfrac{109}{3}x + \dfrac{13}{2} y - \dfrac{34}{3} z\;,\cr
862 & = & 67 x + \dfrac{77}{2}y + 42 z \;,\cr
\dfrac{206}{g_R^2(M_R)}-484
& = & 19 x -16 y + 4 z\;,
\end{eqnarray}
where $x$, $y$, and $z$ are defined as in Eq.~(\ref{xyzDef}).
Solving this system for $x$, $y$, and $z$ we find
\begin{eqnarray}
x & = & \hphantom{-0}2.3 + \dfrac{2.71}{g_R^2(M_R)} \;,\cr
y & = & \hphantom{-}30.2 - \dfrac{8.72}{g_R^2(M_R)} \;,\cr
z & = & -10.8 + \dfrac{3.68}{g_R^2(M_R)} \;.
\end{eqnarray}
Demanding that both $y$ and $z$ be positive restricts $g_R(M_R)$ to the range
\begin{equation}
0.54 \;<\; g_R(M_R) \;<\; 0.58\;.
\end{equation}
The lower bound corresponds to the case considered in Ref.~\citen{Chamseddine:2015ata} at which $y=0$.
Since we would like to minimize $x$, and thereby $M_R$, we set $g_R(M_R)$ to the
upper bound of this range where $x=10.2$, $y=4.6$, and $z=0$. 
This corresponds to
\begin{equation}
M_R \;=\; 1.5\times 10^{12}\,\mathrm{GeV}\;,\qquad
M_C \;=\; M_D \;=\; M_U \;=\; 6\times 10^{14}\,\mathrm{GeV}\;.
\end{equation}
The unified coupling in this case is 
$g_L(M_U)=g_R(M_U)=g_4(M_U)=0.52$.
The running of the couplings for this case is shown in Fig.~\ref{RGrunning3}(b).

\bigskip

Comparing the two cases, allowing $M_R\neq M_C$ has lowered $M_R$ from $10^{13}\,\mathrm{GeV}$
to $10^{12}\,\mathrm{GeV}$.  This is due to the bifurcation of $g_4$ into
$g_3$ and $g_{BL}$ at $M_C$.
The hypercharge coupling at $M_R$ must be matched to $g_R$ and 
$g_{4}$ if $M_R=M_C$, but it will be matched to $g_R$ and $g_{BL}$
if $M_R\neq M_C$.  Since $g_{BL}$ decreases in II, one can allow
$g_R$ to increase further to generate the numerically correct value for $g_1$.
This lowers the scale $M_R$.
However, $10^{12}\,\mathrm{GeV}$ is still too large compared to the TeV scale.
This lowering is also at the expense of $G_{224}$ breaking immediately to $G_{2213}$ as the
model emerges from the underlying NCG theory.

\medskip
\bigskip
\item{Pati-Salam with fundamental Higgs fields}
\bigskip

\begin{table}[t]
\tbl{Higgs content of Model B of Ref.~\citen{Chamseddine:2015ata}.
In Ref.~\citen{Chamseddine:2015ata}, the model emerges with symmetry $G_{224}$ at
$M_U=M_D$.  This breaks directly to $G_{213}$ of the SM at $M_C=M_R$.
We modify this process by allowing $M_C\neq M_R$, inserting energy interval II
with symmetry $G_{2213}$ between intervals III and I.
The Higgs content in interval II is based on the ESH.
The particle content and RG coefficients in intervals I and II are the same as those listed
in Table~\ref{PSa1}.}
{\begin{tabular}{c|l|l}
\hline
$\vphantom{\Big|}$ Interval & Higgs content & RG coefficients \\
\hline
$\vphantom{\Biggl|}$ III
& $\phi(2,2,1),\;H(1,1,6),\;\Delta_R (1,3,10),$ 
& $\left( a_{L},a_{R},a_{4}\right)^\mathrm{III}=\left(2,\dfrac{26}{3},-2\right)$
\\
& $\widetilde{\Sigma}(2,2,15)$
& $\vphantom{\Big|}$ 
\\
\hline
$\vphantom{\Biggl|}$  II 
& $\phi(2,2,0,1),\;\Delta_{R1}(1,3,2,1)$ 
& $\left(a_{L},a_{R},a_{BL},a_3\right)^\mathrm{II}=\left(-3,\dfrac{-7}{3},\dfrac{11}{3},-7\right)$
\\
\hline
$\vphantom{\Biggl|}$ I  
& $\phi_2(2,1,1)$ 
& $\left(a_{1},a_{2},a_{3}\right)^\mathrm{I}=\left(\dfrac{41}{6},\dfrac{-19}{6},-7\right)$ 
\\
\hline
\end{tabular}}
\label{NCG-Model-B}
\end{table}

The Higgs content of Model B of Ref.~\citen{Chamseddine:2015ata}
is shown in Table~\ref{NCG-Model-B}, together with what the
Higgs content in interval II would be under the ESH if the condition $M_C=M_R$
were relaxed.
As in Model A, it is assumed that $M_U=M_D$.
We first follow Ref.~\citen{Chamseddine:2015ata} and also assume $M_C=M_R$
and find 
\begin{eqnarray}
g_R(M_R) & = & 0.48\;,\cr
M_R\;=\;M_C & = & 1.5\times 10^{11}\,\mathrm{GeV}\;,\cr
M_U\;=\;M_D & = & 5.4\times 10^{16}\,\mathrm{GeV}\;.
\end{eqnarray}
The unified coupling is
$g_L(M_U)=g_R(M_U)=g_4(M_U)=0.59$.
The running of the couplings for this case is shown in Fig.~\ref{RGrunning4}(a).

\bigskip

Let us now relax the condition $M_C=M_R$ and insert the energy interval II
with symmetry $G_{2213}$ between intervals I and III.
Without going into detail, we list the Higgs fields that survive via the ESH
into II from III in Table~\ref{NCG-Model-B}.
Note that the Higgs content in I and II are exactly the same as the non-unified 
Pati-Salam model we considered earlier.
In the exact repeat of our analysis of Model A, it can be shown that
for the ordering of the symmetry breaking scale to be maintained, 
$g_R(M_R)$ is restricted to the range
\begin{equation}
0.48\;<\;g_R(M_R)\;<\;0.56\;,
\end{equation}
with the higher bound giving the smallest possible $M_R$.
This is found to be
\begin{equation}
M_R \;=\; 1.1\times 10^{9}\,\mathrm{GeV}\;,\qquad
M_C \;=\; M_D \;=\; M_U \;=\; 4.4\times 10^{16}\,\mathrm{GeV}\;.
\end{equation}
with the unified coupling $g_L(M_U)=g_R(M_U)=g_4(M_U)=0.52$.
The running of the couplings for this case is shown in Fig.~\ref{RGrunning4}(b).

\begin{figure}[t]
\subfigure[$g_R(M_R)=0.59$]{\includegraphics[width=6.5cm]{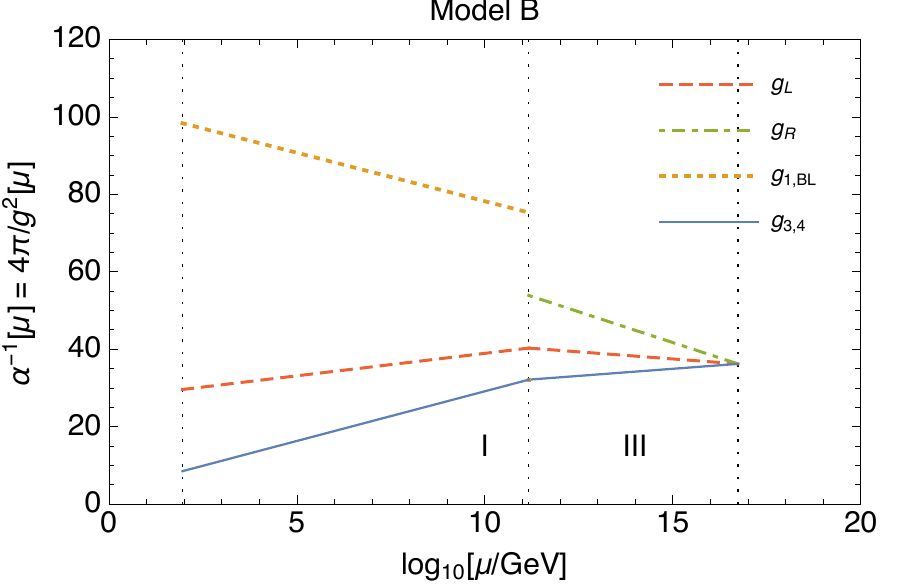}}\subfigure[$g_R(M_R)=0.52$]{\includegraphics[width=6.5cm]{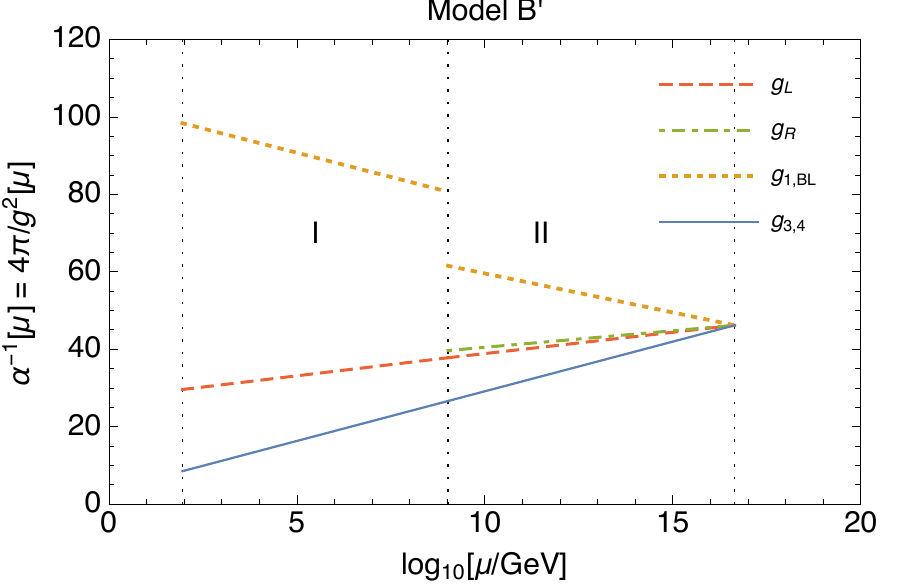}}
\caption{Running of the gauge couplings for Model B of Ref.~\citen{Chamseddine:2015ata}.
with (a) $G_{224}$ breaking directly into $G_{213}$, and 
(b) $G_{224}$ breaks immediately to $G_{2213}$ as it emerges.
}
\label{RGrunning4}
\end{figure}

\bigskip

While this result is somewhat more promising than Model A, $M_R$ is still to large,
as is the value of $g_R(M_R)$ necessary for $M_R$ to be pushed down to this scale.
Let us see if the situation may be improved by relaxing the ESH as we did for the
non-unified Pati-Salam case.  We will allow some or all of the colored $\Delta_R$ fields
to survive into interval II to enhance the difference between $g_L$ and $g_R$.
We consider the same three cases listed in Table~\ref{PSa2}.

\renewcommand{\theenumii}{\roman{enumii}}
\begin{enumerate}
\item $\Delta_{R1}$ and $\Delta_{R3}$ survive:

To maintain the ordering of the symmetry breaking scales, it is found that
$g_R(M_R)$ is restricted to the narrow range
\begin{equation}
0.48\;<\; g_R(M_R) \;<\; 0.51\;.
\end{equation}
As $g_R(M_R)$ is increased, $M_R/M_Z$ and $M_D/M_C$ decrease while $M_C/M_R$ increases.
In terms of scale, $M_R$ decreases while both $M_C$ and $M_D$ increase.
The upper bound of this range is when $M_R/M_Z=1$, 
so this case actually allows for $M_R=5\,\mathrm{TeV}$.
The other parameters in this case is
\begin{eqnarray}
g_R(M_R) & = & 0.51\;,\cr
M_R & = & 5\times 10^3\,\mathrm{GeV}\;,\cr
M_C & = & 5\times 10^{15}\,\mathrm{GeV}\;,\cr
M_U\;=\;M_D & = & 8\times 10^{17}\,\mathrm{GeV}\;,
\end{eqnarray}
with the unified coupling $g_L(M_U)=g_R(M_U)=g_4(M_U)=0.54$.
The running of the couplings for this case is shown in Fig.~\ref{RGrunning5}(a).

\bigskip
\item $\Delta_{R1}$ and $\Delta_{R6}$ survive:

To maintain the ordering of the symmetry breaking scales, it is found that
$g_R(M_R)$ is restricted to the range
\begin{equation}
0.42\;<\; g_R(M_R) \;<\; 0.48\;,
\end{equation}
with smaller $g_R(M_R)$ associated with smaller $M_R$, which drops down to $M_Z$ 
at the lower bound.  Imposing $M_R=5\,\mathrm{TeV}$ we obtain:
\begin{eqnarray}
g_R(M_R) & = & 0.43\;,\cr
M_R & = & 5\times 10^3\,\mathrm{GeV}\;,\cr
M_C & = & 2\times 10^{10}\,\mathrm{GeV}\;,\cr
M_U\;=\;M_D & = & 3\times 10^{20}\,\mathrm{GeV}\;,
\end{eqnarray}
with the unified coupling $g_L(M_U)=g_R(M_U)=g_4(M_U)=0.63$.
The running of the couplings for this case is shown in Fig.~\ref{RGrunning5}(b).
Maintaining $M_D$ below $10^{19}$ requires
\begin{equation}
0.45 \;<\; g_R(M_R)\;.
\end{equation}
Selecting this boundary value for $g_R(M_R)$, we find
\begin{eqnarray}
g_R(M_R) & = & 0.45\;,\cr
M_R & = & 5\times 10^6\,\mathrm{GeV}\;,\cr
M_C & = & 4\times 10^{10}\,\mathrm{GeV}\;,\cr
M_U\;=\;M_D & = & 1\times 10^{19}\,\mathrm{GeV}\;,
\end{eqnarray}
with the unified coupling $g_L(M_U)=g_R(M_U)=g_4(M_U)=0.62$.
The running of the couplings for this case is shown in Fig.~\ref{RGrunning5}(c).

\begin{figure}[t]
\subfigure[$g_R(M_R)=0.51$]{\includegraphics[width=6.5cm]{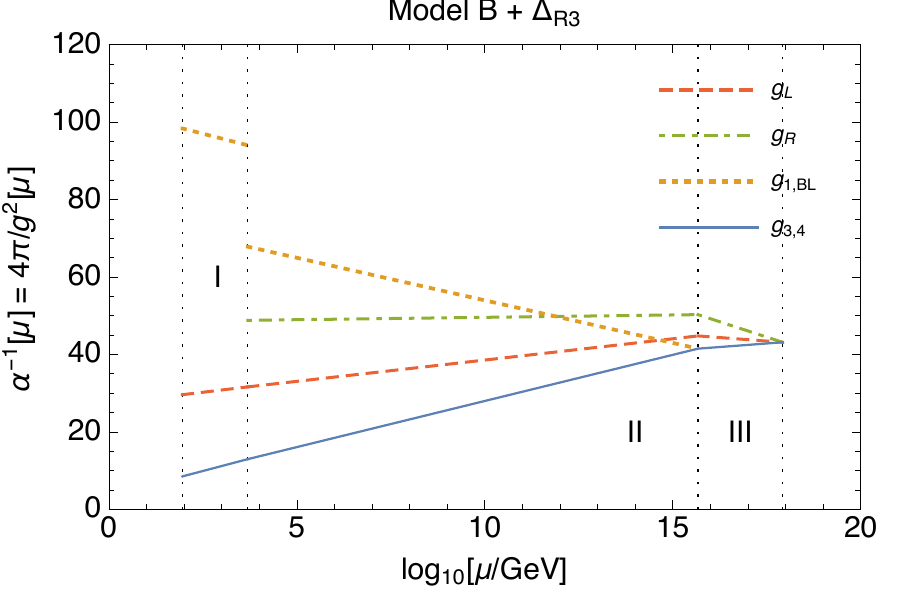}}
\subfigure[$g_R(M_R)=0.43$]{\includegraphics[width=6.5cm]{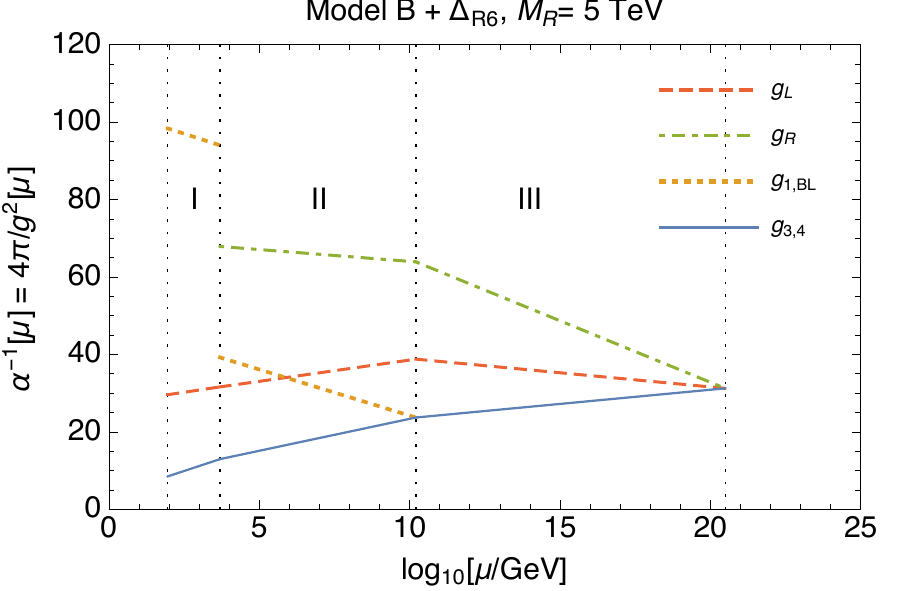}}\subfigure[$g_R(M_R)=0.45$]{\includegraphics[width=6.5cm]{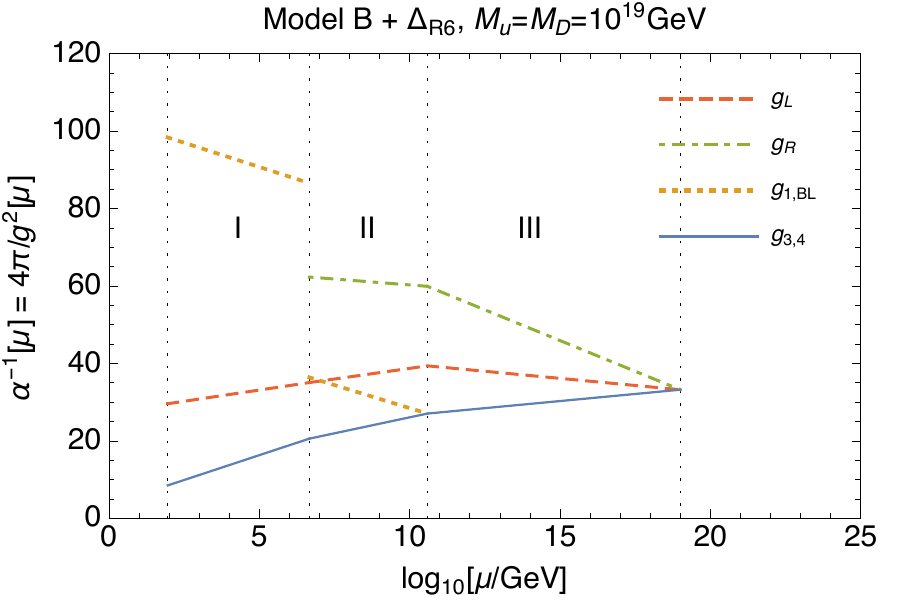}}
\subfigure[$g_R(M_R)=0.42$]{\includegraphics[width=6.5cm]{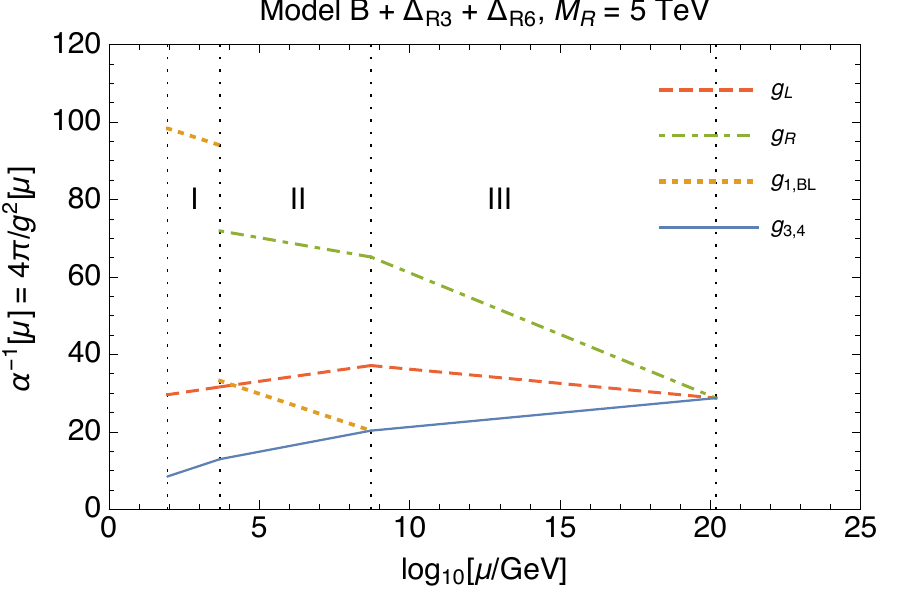}}\subfigure[$g_R(M_R)=0.44$]{\includegraphics[width=6.5cm]{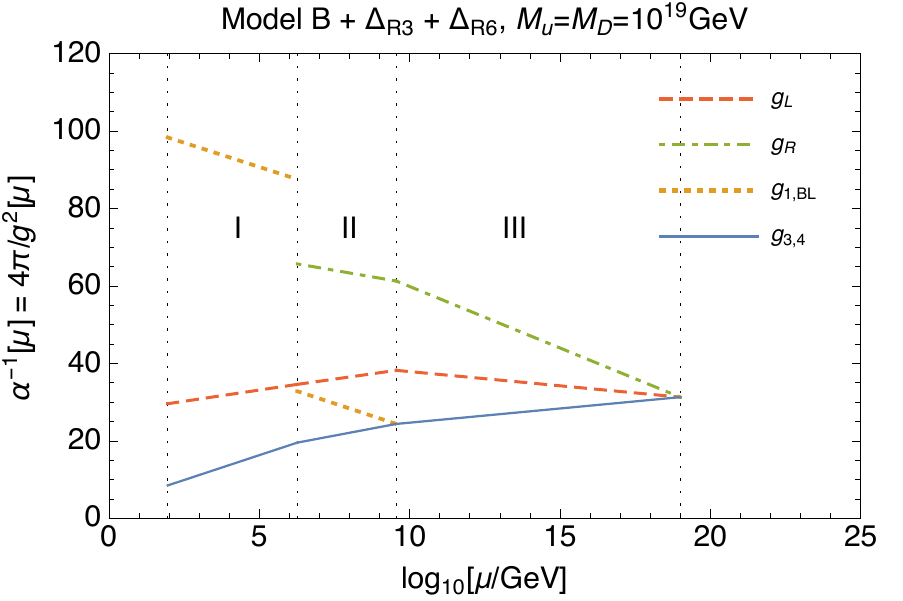}}
\caption{Running of the gauge couplings for Model B of Ref.~\citen{Chamseddine:2015ata}
with extended Higgs content in energy interval II.
In addition to $\Delta_{R1}$, the field $\Delta_{R3}$ survives into II in (a), (d), and (e),
while the field $\Delta_{R6}$ also survives into II in (b), (c), (d) and (e).
In (a), (b), and (d) we impose $M_R=5\,\mathrm{TeV}$.
In (c) and (e) we impose $M_U=M_D=10^{19}\,\mathrm{GeV}$.
}
\label{RGrunning5}
\end{figure}

\bigskip
\item $\Delta_{R1}$, $\Delta_{R3}$, and $\Delta_{R6}$ all survive:

To maintain the ordering of the symmetry breaking scales, it is found that
$g_R(M_R)$ is restricted to the range
\begin{equation}
0.41\;<\; g_R(M_R) \;<\; 0.48\;,
\end{equation}
while to maintain $M_U=M_D$ below $10^{19}\,\mathrm{GeV}$ we must have
\begin{equation}
0.44\;<\; g_R(M_R)\;.
\end{equation}
If we demand $M_R=5\,\mathrm{TeV}$, we find
\begin{eqnarray}
g_R(M_R) & = & 0.42\;,\cr
M_R & = & 5\times 10^3\,\mathrm{GeV}\;,\cr
M_C & = & 5\times 10^{8}\,\mathrm{GeV}\;,\cr
M_U\;=\;M_D & = & 2\times 10^{20}\,\mathrm{GeV}\;,
\end{eqnarray}
with the unified coupling $g_L(M_U)=g_R(M_U)=g_4(M_U)=0.66$.
The running of the couplings for this case is shown in Fig.~\ref{RGrunning5}(d).
If we demand $M_U=M_D=10^{19}\,\mathrm{GeV}$, we find
\begin{eqnarray}
g_R(M_R) & = & 0.44\;,\cr
M_R & = & 2\times 10^6\,\mathrm{GeV}\;,\cr
M_C & = & 4\times 10^{9}\,\mathrm{GeV}\;,\cr
M_U\;=\;M_D & = & 1\times 10^{19}\,\mathrm{GeV}\;,
\end{eqnarray}
with the unified coupling $g_L(M_U)=g_R(M_U)=g_4(M_U)=0.63$.
The running of the couplings for this case is shown in Fig.~\ref{RGrunning5}(e).

\end{enumerate}

\medskip
\bigskip
\item{Left-right symmetric Pati-Salam with fundamental Higgs fields}
\smallskip

\begin{table}[t]
\tbl{Higgs content of Model C of Ref.~\citen{Chamseddine:2015ata}.
In Ref.~\citen{Chamseddine:2015ata}, the model emerges with symmetry $G_{224D}$ at
$M_U$.  This breaks directly to $G_{213}$ of the SM at $M_D=M_C=M_R$.
We modify this process by allowing $M_D\neq M_C\neq M_R$, inserting energy intervals II
and III with symmetries $G_{2213}$ and $G_{224}$, respectively, between intervals I and IV.
The Higgs content in intervals I, II, and III are based on the ESH.
An extra D-parity singlet field $\sigma(1,1,1)$ is introduced in interval IV to break parity spontaneously.
The particle content and RG coefficients in intervals I and II are the same as those listed
in Table~\ref{PSa1}.}
{\begin{tabular}{c|l|l}
\hline
$\vphantom{\Big|}$ Interval & Higgs content & RG coefficients \\
\hline
$\vphantom{\Bigg|}$ IV 
& $\phi(2,2,1),\;H(1,1,6)\times 2,\;\widetilde{\Sigma}(2,2,15)$ 
& $\left( a_{L},a_{R},a_{4}\right)^\mathrm{IV}
 =\left(\dfrac{26}{3},\dfrac{26}{3},\dfrac{4}{3}\right)$ 
\\
& $\Delta_R(1,3,10),\;\Delta_L(3,1,10),\;\sigma(1,1,1)$ & \\
& & \\
\hline
$\vphantom{\Bigg|}$ III 
& $\phi(2,2,1),\;H(1,1,6),\;\Delta_R(1,3,10)$ 
& $\left( a_{L},a_{R},a_{4}\right)^\mathrm{III}
 =\left(-3,\dfrac{11}{3},\dfrac{-22}{3}\right)$ \\
\hline
$\vphantom{\Bigg|}$ II 
& $\phi(2,2,0,1),\;\Delta_{R1}(1,3,2,1)$ 
& $\left(a_{L},a_{R},a_{BL},a_3\right)^\mathrm{II}=\left(-3,\dfrac{-7}{3},\dfrac{11}{3},-7\right)$
\\
\hline
$\vphantom{\Bigg|}$ I  
& $\phi_2(2,1,1)$ 
& $\left(a_{1},a_{2},a_{3}\right)=\left(\dfrac{41}{6},\dfrac{-19}{6},-7\right)$ 
\\
\hline
\end{tabular}}
\label{NCG-Model-C}
\end{table}

Finally, the last and most general scenario of Ref.~\citen{Chamseddine:2015ata}
is where $G_{224D}$ instead of $G_{224}$ is the emergent symmetry of the spectral action. 
The assumed Higgs content of the model is shown in Table~\ref{NCG-Model-C}.

First, assuming $M_D=M_C=M_R$ as in Ref.~\citen{Chamseddine:2015ata}, we solve
Eqs.~(\ref{NCGeq1}) through (\ref{NCGeq3}) and find
\begin{eqnarray}
g_R(M_R) & = & 0.54\;,\cr
M_D\;=\;M_C\;=\; M_R & = & 5\times 10^{13}\,\mathrm{GeV}\;,\cr
M_U & = & 3\times 10^{15}\,\mathrm{GeV}\;,
\end{eqnarray}
with the unified coupling $g_L(M_U)=g_R(M_U)=g_4(M_U)=0.58$.
The running of the couplings for this case is shown in Fig.~\ref{RGrunning6}(a).

\begin{figure}[t]
\subfigure[$g_R(M_R)=0.51$]{\includegraphics[width=6.5cm]{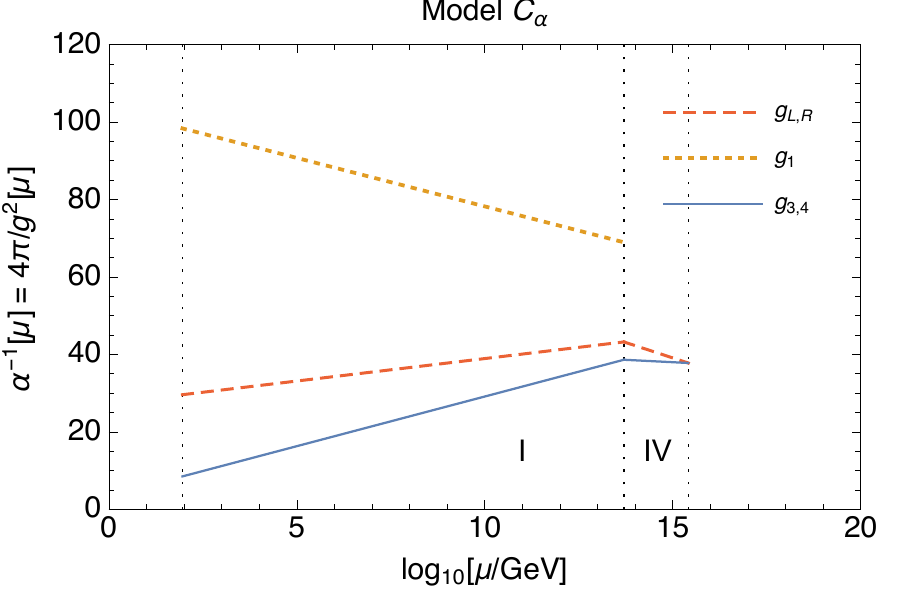}}\subfigure[$g_R(M_R)=0.56$]{\includegraphics[width=6.5cm]{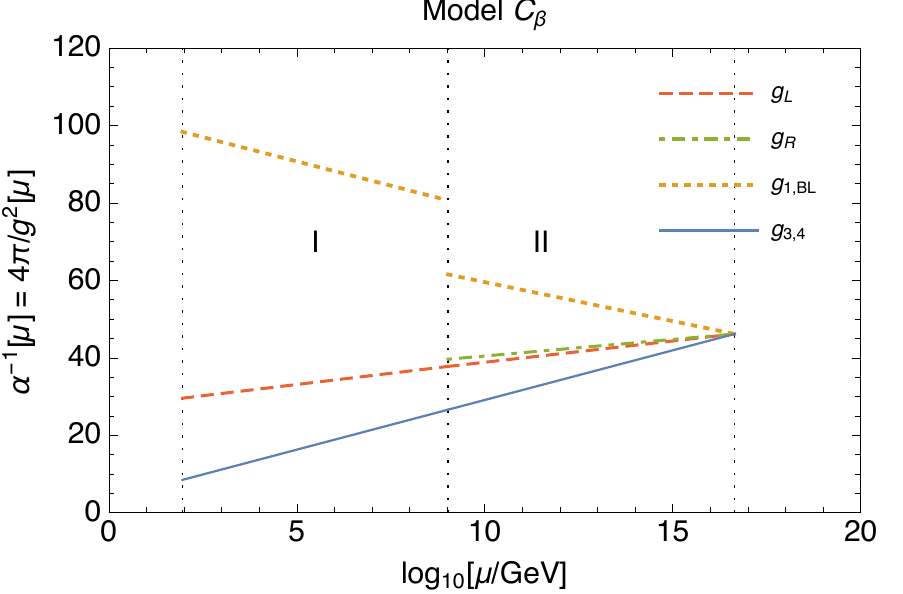}}
\subfigure[$g_R(M_R)=0.49$]{\includegraphics[width=6.5cm]{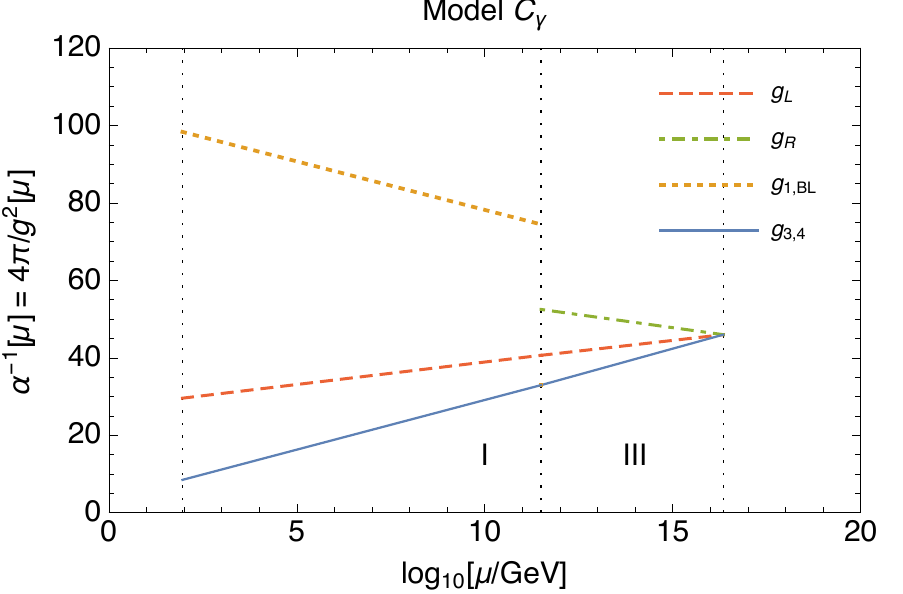}}\subfigure[$g_R(M_R)=0.51$]{\includegraphics[width=6.5cm]{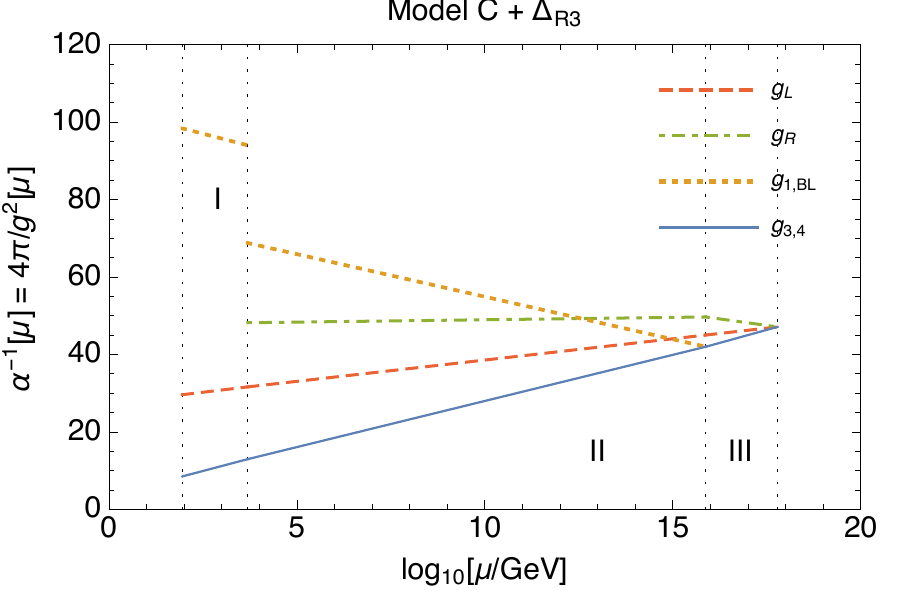}}
\subfigure[$g_R(M_R)=0.45$]{\includegraphics[width=6.5cm]{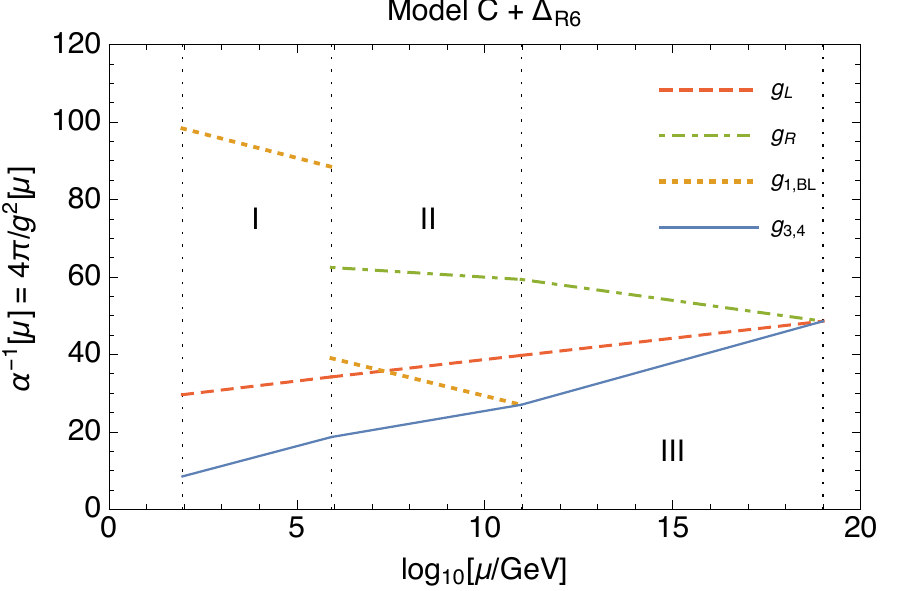}}\subfigure[$g_R(M_R)=0.43$]{\includegraphics[width=6.5cm]{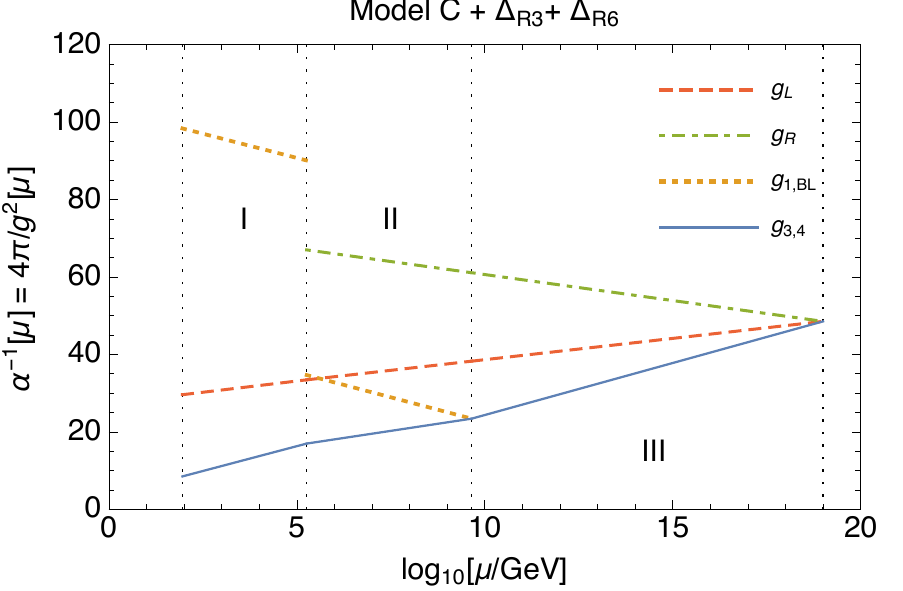}}
\caption{(a) Running of the gauge couplings for Model C of Ref.~\citen{Chamseddine:2015ata}
where $M_D=M_C=M_R$,
(b) $M_U=M_D=M_C$, 
(c) $M_U=M_D$, $M_C=M_R$,
(d) with $\Delta_{R3}$ surviving in II,
(e) with $\Delta_{R6}$ surviving in II, and
(f) with $\Delta_{R3}$ and $\Delta_{R6}$ surviving in II.
}
\label{RGrunning6}
\end{figure}

\begin{figure}[t]
\subfigure[]{\includegraphics[width=6cm]{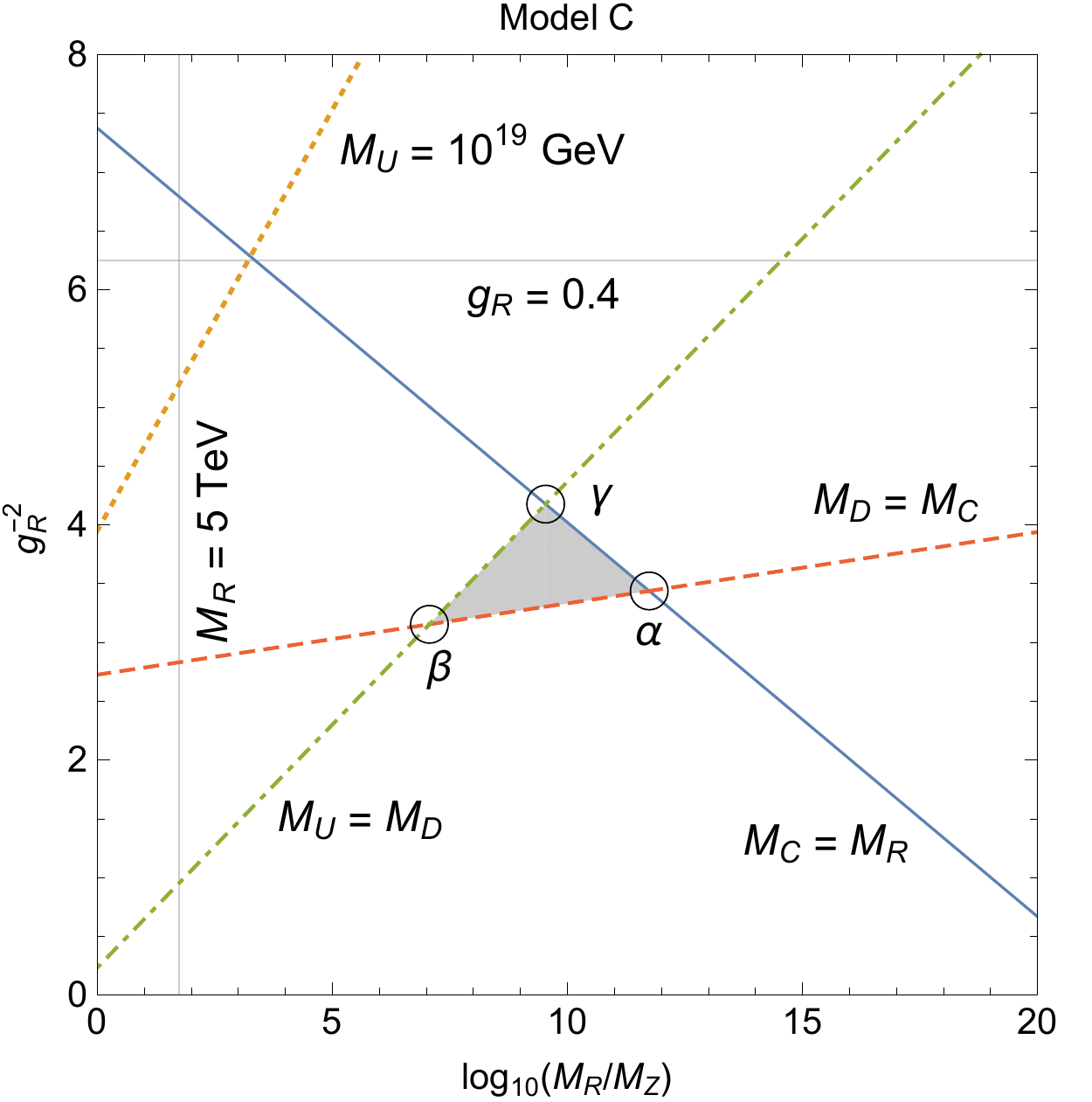}}\subfigure[]{\includegraphics[width=6cm]{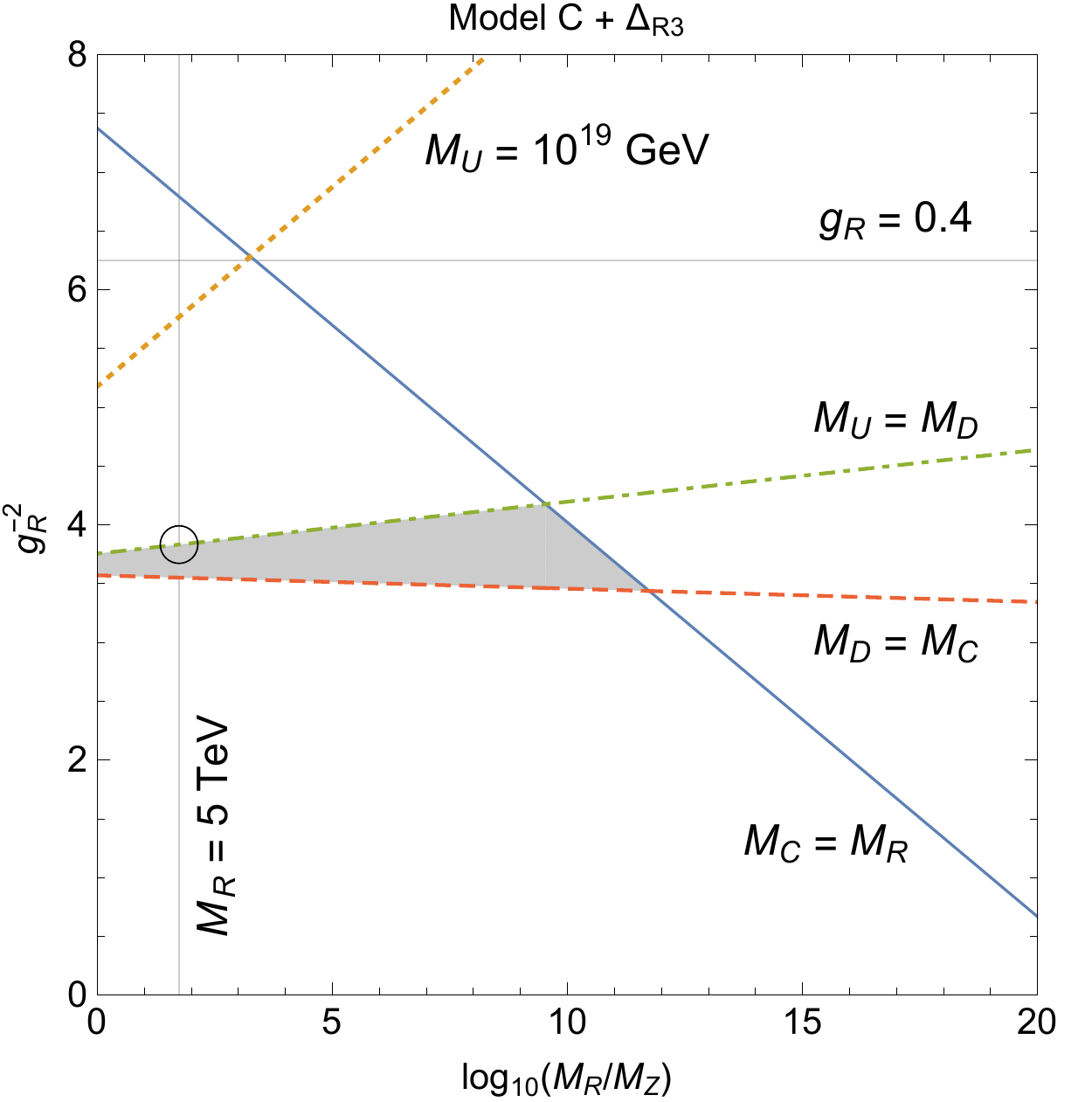}}
\subfigure[]{\includegraphics[width=6cm]{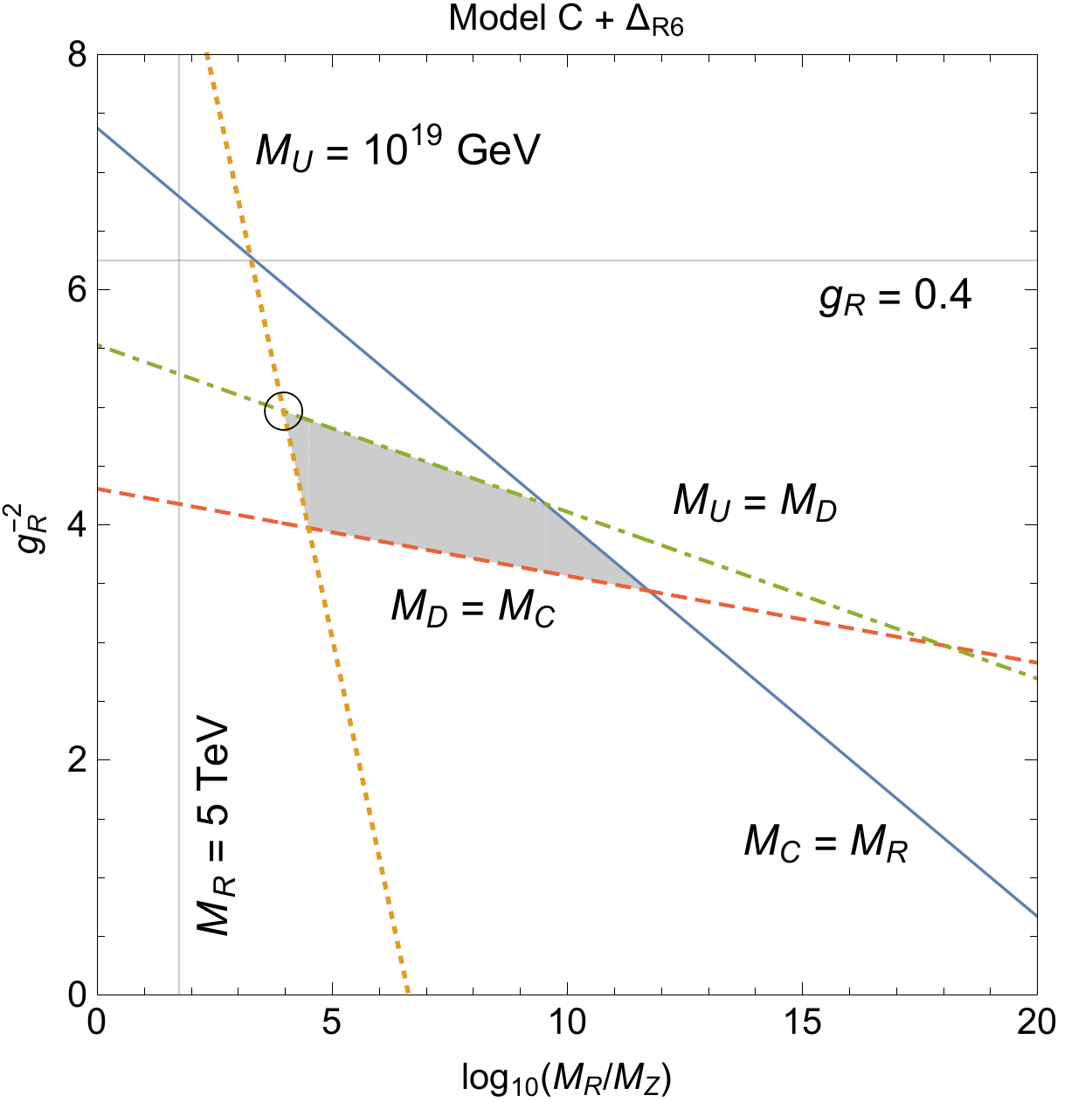}}\subfigure[]{\includegraphics[width=6cm]{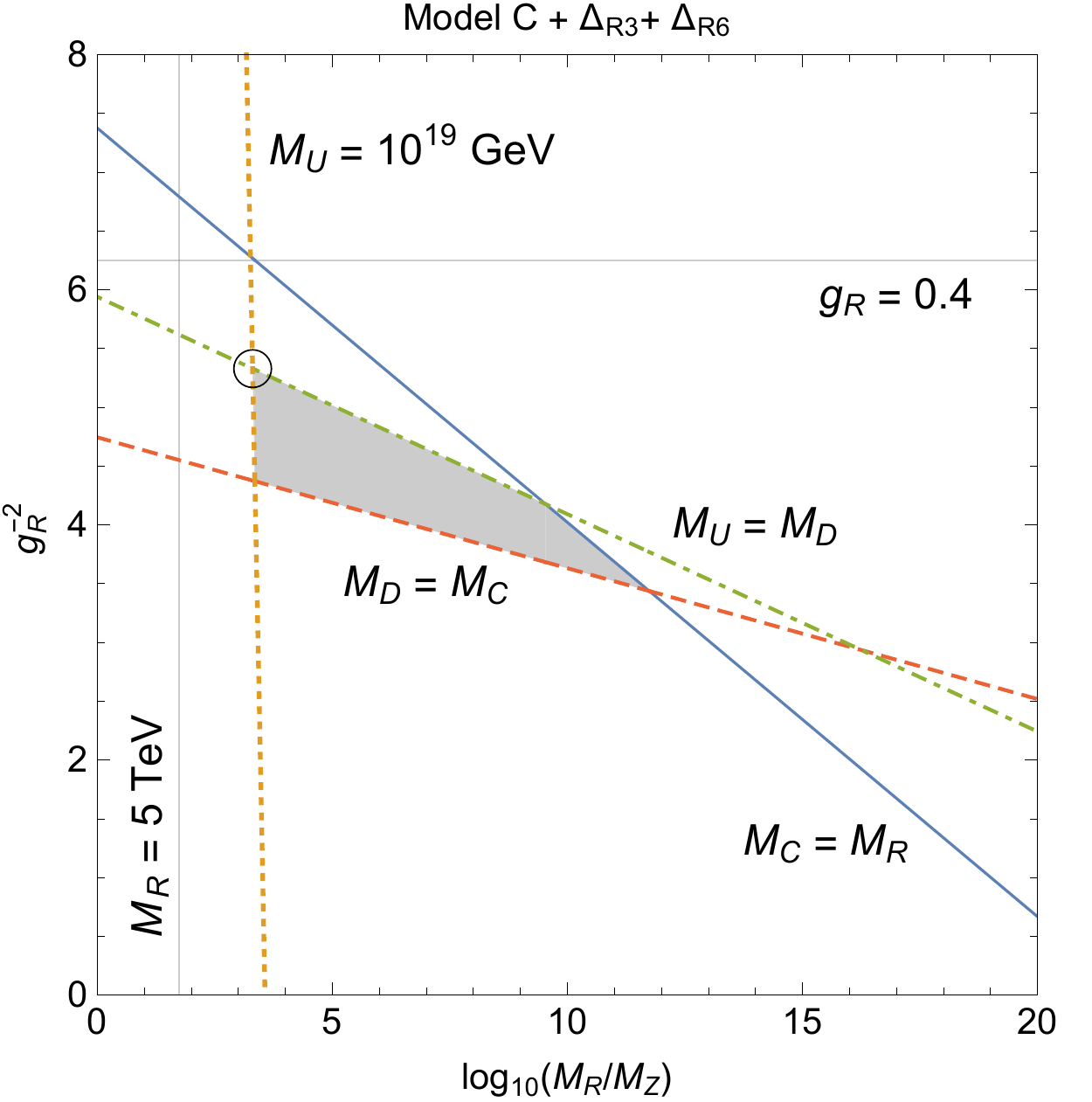}}
\caption{
(a) For Model C, the values of $x=\log_{10}M_R/M_Z$ and $g^{-2}_R(M_R)$ must lie inside the shaded
triangle shown to maintain the ordering of the symmetry breaking scales.
Ref.~\citen{Chamseddine:2015ata} selects the values at point $\alpha$, where $M_D=M_C=M_R$.
$M_R$ is minimized at point $\beta$ where $M_U=M_D=M_C$, 
while $g_R(M_R)$ is minimized at point $\gamma$ where $M_U=M_D$ and $M_C=M_R$.
(b), (c), and (d) show how the allowed region changes with the addition 
of extra colored $\Delta_R$ fields in energy interval II.
The requirement that $M_U\le 10^{19}\,\mathrm{GeV}$ demands that one stay to the right of the dotted line, and this restricts us to the interiors of the
shaded quadrangles shown.
Consequently, only case (b) allows for $M_R=5\,\mathrm{GeV}$.
In all three cases, $g_R(M_R)$ is minimized for
a given choice of $M_R$ when $M_U=M_D$. 
The optimum points for each case discussed in the text are indicated by circles.
}
\label{xgRallowed}
\end{figure}

\bigskip

We next relax the relation $M_D=M_C=M_R$ and insert energy intervals II and III
in between intervals I and IV with the Higgs content listed in Table~\ref{NCG-Model-C}.
Eqs.~(\ref{NCGeq1}) through (\ref{NCGeq3}) now read
\begin{eqnarray}
401 & = & \dfrac{109}{3}x + 19 y + \dfrac{34}{3} z - \dfrac{44}{3} w\;,\vphantom{\Bigg|}\cr
862 & = & 67 x + 51 y + 46 z + 44 w \;,\cr
\dfrac{206}{g_R^2(M_R)}-484
& = & 19 x + 4 y + 40 z\;,
\end{eqnarray}
where $x$, $y$, and $z$ are defined as in Eq.~(\ref{xyzDef}), and
$w=\log_{10}M_U/M_D$.
Solving this system for $y$, $z$, and $w$ we find:
\begin{eqnarray}
y & = & 30.3 -1.38 x -\dfrac{4.11}{g_R^2(M_R)}\;,\cr
z & = & 15.1 + 0.34 x -\dfrac{5.56}{g_R^2(M_R)}\;,\cr
w & = & 0.24 + 0.43 x -\dfrac{1.04}{g_R^2(M_R)}\;.
\end{eqnarray}
Demanding that $y$, $z$, and $w$ are all positive restricts $x=\log_{10}M_R/M_Z$ and
$g_R^{-2}(M_R)$ to the triangular region shown in Fig.~\ref{xgRallowed}(a).
It is clear from the figure that $M_R$ is minimized when $M_U=M_D=M_C$, that is,
energy regions III and IV are collapsed and only I and II remain.
On the other hand, $g_R(M_R)$ is minimized when $M_U=M_D$ and $M_C=M_R$, that is,
energy regions II and IV are collapsed and only I and III remain.

\bigskip
For the $M_U=M_D=M_C$ case, we find
\begin{eqnarray}
g_R(M_R) & = & 0.56\;,\cr
M_R & = & 1\times 10^{9}\,\mathrm{GeV}\;,\cr
M_U\;=\;M_D\;=\;M_C & = & 4\times 10^{16}\,\mathrm{GeV}\;,
\end{eqnarray}
with the unified coupling $g_L(M_U)=g_R(M_U)=g_4(M_U)=0.52$.
The running of the couplings for this case is shown in Fig.~\ref{RGrunning6}(b).

\bigskip
For the $M_U=M_D$, $M_C=M_R$ case, we find
\begin{eqnarray}
g_R(M_R) & = & 0.49\;,\cr
M_C\;=\;M_R & = & 3\times 10^{11}\,\mathrm{GeV}\;,\cr
M_U\;=\;M_D & = & 2\times 10^{16}\,\mathrm{GeV}\;,
\end{eqnarray}
with the unified coupling $g_L(M_U)=g_R(M_U)=g_4(M_U)=0.52$.
The running of the couplings for this case is shown in Fig.~\ref{RGrunning6}(c).

\bigskip

Again, the values of $M_R$ and $g_R(M_R)$ thus obtained
are more promising that what could be achieved in Model A, but nevertheless they are both
still too large.
So let us relax the ESH in energy interval II again to see whether things are improved.
As we did for Model B, we consider the three cases listed in Table~\ref{PSa2}.
The allowed regions in $(x,g_R^{-2})$ space is shown in Figs.~\ref{xgRallowed}(b) through \ref{xgRallowed}(d).
Taking $M_R$ to be as close to $5\,\mathrm{TeV}$ as possible while minimizing $g_R(M_R)$ and maintaining $M_U\le 10^{19}\,\mathrm{GeV}$ leads to the following optimum solutions:

\begin{enumerate}
\item ${\Delta_{1R}}$ and ${\Delta_{3R}}$ survive:
\begin{eqnarray}
g_R(M_R) & = & 0.51\;,\cr
M_R & = & 5\times 10^{3}\,\mathrm{GeV}\;,\cr
M_C & = & 8\times 10^{15}\,\mathrm{GeV}\;,\cr
M_U\;=\;M_D & = & 6\times 10^{17}\,\mathrm{GeV}\;,
\end{eqnarray}
with the unified coupling $g_L(M_U)=g_R(M_U)=g_4(M_U)=0.52$.
The running of the couplings for this case is shown in Fig.~\ref{RGrunning6}(d).

\bigskip
\item ${\Delta_{1R}}$ and ${\Delta_{6R}}$ survive:
\begin{eqnarray}
g_R(M_R) & = & 0.45\;,\cr
M_R & = & 8\times 10^{5}\,\mathrm{GeV}\;,\cr
M_C & = & 9\times 10^{10}\,\mathrm{GeV}\;,\cr
M_U\;=\;M_D & = & 1\times 10^{19}\,\mathrm{GeV}\;,
\end{eqnarray}
with the unified coupling $g_L(M_U)=g_R(M_U)=g_4(M_U)=0.51$.
The running of the couplings for this case is shown in Fig.~\ref{RGrunning6}(e).

\bigskip
\item ${\Delta_{1R}}$, ${\Delta_{3R}}$, and ${\Delta_{6R}}$ survive:
\begin{eqnarray}
g_R(M_R) & = & 0.43\;,\cr
M_R & = & 2\times 10^{5}\,\mathrm{GeV}\;,\cr
M_C & = & 4\times 10^{9}\,\mathrm{GeV}\;,\cr
M_U\;=\;M_D & = & 1\times 10^{19}\,\mathrm{GeV}\;,
\end{eqnarray}
with the unified coupling $g_L(M_U)=g_R(M_U)=g_4(M_U)=0.51$.
The running of the couplings for this case is shown in Fig.~\ref{RGrunning6}(f).

\end{enumerate}

\end{enumerate}

\subsection{Summary of Results}

In this section, we have looked at whether the IR conditions $M_R=5\,\mathrm{TeV}$
and $g_R(M_R)=0.4$ could be realized within left-right symmetric, and unified left-right 
symmetric Pati-Salam models in which the unification/emergence scale is below the Planck mass. 
The left-right symmetric Pati-Salam demands the unification
of $g_L$ and $g_R$, while the unified left-right symmetric Pati-Salam demands further
unification of $g_L$ and $g_R$ with $g_4$.
The requirements that these couplings unify at a single scale, and the matching conditions
between $g_1$, $g_{BL}$, and $g_R$ at $M_R$, and that between $g_{BL}$, $g_3$ and $g_4$ 
at $M_C$, place conflicting demands on the various symmetry breaking scales, and it is
found that the target IR conditions cannot be realized so easily.
In particular, if the Higgs content at various energy intervals is determined
based on the Extended Survival Hypothesis (ESM), $M_R$ and $g_R(M_R)$ tend to be much larger 
than our target values.
Lowering these values requires the breaking of ESH.
The most promising cases are Models B and C of Ref.~\citen{Chamseddine:2015ata} 
with the colored $\Delta_{3R}$ field surviving below $M_C$.
We note that this may put the $\Delta_{3R}$ particles within reach of the LHC. 
But even for those cases $g_R(M_R)$ cannot be made as low as $0.4$.
In all cases, the optimum conditions for minimum $M_R$ and/or minimum $g_R(M_R)$ requires
degeneracies of some of the symmetry breaking scales.

\section{Discussion and Conclusions}

\bigskip

In this paper, we have initiated a purely phenomenological analysis of Connes' 
NCG approach to the SM and beyond, in the light of the latest experimental results from the LHC.  
In particular, we have concentrated on the remarkable left-right symmetric 
structure that is inherent in the NCG of the SM, 
embodied in the unified left-right symmetric Pati-Salam models of Ref.~\citen{Chamseddine:2015ata}, 
and explored its phenomenological consequences by concentrating on the possible existence of a TeV scale $W_R$ boson.
We find that generating a TeV scale $W_R$ boson with the small coupling of $g_R=0.4$ within
NCG motivated models is not trivial and places strong constraints on the particle content
and symmetry breaking scales.

We note that we have also conducted a preliminary analysis of the constraints imposed by proton stability\cite{Mohapatra:1980qe}, 
the $\Delta B=2$ neutron-antineutron and hydrogen-antihydrogen oscillations\cite{Mohapatra:1982aq}
as well as the constraints coming from the inflationary cosmological models\cite{Bertolini:2009qj}. 
In principle, these constraints are not prohibitive of the phenomenological analysis carried out in this paper.

While our analysis could suggest that the NCG motivated unified left-right Pati-Salam model is not favored phenomenologically by the current LHC data, we note the possibility that the current approach of grafting the NCG spectral action to RG evolution of standard QFT at the GUT scale may not capture the true nature and predictions of NCG theories.

Finally, we address the closely related question of the hierarchy problem.
One of the most interesting aspects of the NCG of the SM and its Pati-Salam-like completion is the existence of the GUT scale which can be found in the close proximity to the Planck scale, i.e. the scale of quantum gravity. 
Given this fact as well as the presence of a hidden fundamental non-commutative structure in this approach, this suggests that the hierarchy problem should get a quantum gravitational, and not an effective field theory treatment. 
The more convincing physical meaning of this GUT scale also comes after one realizes that Connes' approach also produces a gravity sector in parallel with the standard model (and its Pati-Salam completion) and thus the GUT scale should be viewed as being close to the natural scale of gravity, i.e. Planck scale, and indeed the two scales are not that far apart in the non-commutative approach.
In particular if one views quantum gravity as having origins in metastring theory \cite{Freidel:2013zga,Freidel:2014qna,Freidel:2015pka} then one finds the fundamental non-commutative structure, and also, the two-scale
renormalization group, which sheds new light on such fundamental issues as the hierarchy problem: the two scales that feature in Refs.~\citen{Freidel:2013zga,Freidel:2014qna,Freidel:2015pka} are both the UV and the IR scales and thus the
stability of the Higgs mass 
becomes two-fold, both with respect to the UV and to the IR.
In other words, the question is now not only why the Higgs mass is not of the Planck scale (or the GUT scale) but also why the Higgs mass is not of the Hubble (vacuum energy) scale.
It is well known that numerologically, the Higgs scale ($\sim 1\,\mathrm{TeV}$) is the geometric mean between these two scales, at the point of a UV/IR invariant energy scale. The Higgs scale also naturally appears as a geometric scale in Connes' non-commutative geometry approach, in complete
analogy with the geometric meaning of the Planck and the Hubble scales. Actually, because of the appearance of gravity and the standard model Lagrangians in the Connes's spectral action, and because of the discrete nature of the Higgs dimension, there is a natural Higgs-like degree of freedom on the gravity side -- a Brans-Dicki-Jordan-like scalar -- which can be argued to contribute to the geometric warping of the Higgs discrete dimension. This is similar to
the infinite extra dimensional scenarios, however, without infinite extra dimensions. \cite{connes-book2, Chamseddine:2010ud}

In our view the approach based on NCG (and its related proposal based on the superconnection approach \cite{Aydemir:2013zua,Aydemir:2014ama}) offers a new and, phenomenologically, almost completely unexplored view on the rationale for the SM and also for its natural completion. This approach also offers a possibly exciting relation with the fundamental physics of quantum gravity, thus relating the infrared physics of the current exciting experimental searches conducted at the LHC to the hidden ultraviolet physics of quantum theory of space and time.


\section*{Acknowledgments}

We would like to thank L.~Boyle, A.~H.~Chamseddine, L.~Freidel, R.~Leigh, and A.~Shapere for helpful communications and stimulating conversations.
UA is supported by the Swedish Research Council under contract 621-2011-5107 and DM was supported in part by the U.S. Department of Energy, grant DE-FG02-13ER41917, task A.

\appendix

\section{Derivation of Relations Between Symmetry Breaking Scales}

\begin{eqnarray}
\dfrac{1}{g_1^2(M_Z)}
& = & \dfrac{1}{g_1^2(M_R)} + \dfrac{a_1^\mathrm{I}}{8\pi^2}\ln\dfrac{M_R}{M_Z} \cr
& = & \dfrac{1}{g_R^2(M_R)} + \dfrac{1}{g_{BL}^2(M_R)} + \dfrac{a_1^\mathrm{I}}{8\pi^2}\ln\dfrac{M_R}{M_Z} \cr
& = & \dfrac{1}{g_R^2(M_C)} + \dfrac{1}{g_{BL}^2(M_C)} 
+ \dfrac{\left(a_R + a_{BL}\right)^\mathrm{II}}{8\pi^2}\ln\dfrac{M_C}{M_R}
+ \dfrac{a_1^\mathrm{I}}{8\pi^2}\ln\dfrac{M_R}{M_Z} \cr
& = & \dfrac{1}{g_R^2(M_C)} + \dfrac{2}{3}\,\dfrac{1}{g_4^2(M_C)}
+ \dfrac{\left(a_R + a_{BL}\right)^\mathrm{II}}{8\pi^2}\ln\dfrac{M_C}{M_R}
+ \dfrac{a_1^\mathrm{I}}{8\pi^2}\ln\dfrac{M_R}{M_Z} \cr
& = & \dfrac{1}{g_R^2(M_D)} + \dfrac{2}{3}\,\dfrac{1}{g_4^2(M_D)}
+ \dfrac{1}{8\pi^2}\left(a_R + \dfrac{2}{3}a_4\right)^\mathrm{III}\ln\dfrac{M_D}{M_C}
\cr & & 
+ \dfrac{\left(a_R + a_{BL}\right)^\mathrm{II}}{8\pi^2}\ln\dfrac{M_C}{M_R}
+ \dfrac{a_1^\mathrm{I}}{8\pi^2}\ln\dfrac{M_R}{M_Z} \cr
& = & \dfrac{1}{g_R^2(M_U)} + \dfrac{2}{3}\,\dfrac{1}{g_4^2(M_U)}
+ \dfrac{1}{8\pi^2}\left(a_R + \dfrac{2}{3}a_4\right)^\mathrm{IV}\ln\dfrac{M_U}{M_D}
\cr & & 
+ \dfrac{1}{8\pi^2}\left(a_R + \dfrac{2}{3}a_4\right)^\mathrm{III}\ln\dfrac{M_D}{M_C}
+ \dfrac{\left(a_R + a_{BL}\right)^\mathrm{II}}{8\pi^2}\ln\dfrac{M_C}{M_R}
+ \dfrac{a_1^\mathrm{I}}{8\pi^2}\ln\dfrac{M_R}{M_Z} 
\;,
\cr
\dfrac{1}{g_2^2(M_Z)}
& = & \dfrac{1}{g_2^2(M_R)} + \dfrac{a_2^\mathrm{I}}{8\pi^2}\ln\dfrac{M_R}{M_Z} \cr
& = & \dfrac{1}{g_L^2(M_R)} + \dfrac{a_2^\mathrm{I}}{8\pi^2}\ln\dfrac{M_R}{M_Z} \cr
& = & \dfrac{1}{g_L^2(M_C)} + \dfrac{a_L^\mathrm{II}}{8\pi^2}\ln\dfrac{M_C}{M_R}
+ \dfrac{a_2^\mathrm{I}}{8\pi^2}\ln\dfrac{M_R}{M_Z} \cr
& = & \dfrac{1}{g_L^2(M_D)} 
+ \dfrac{a_L^\mathrm{III}}{8\pi^2}\ln\dfrac{M_D}{M_C}
+ \dfrac{a_L^\mathrm{II}}{8\pi^2}\ln\dfrac{M_C}{M_R}
+ \dfrac{a_2^\mathrm{I}}{8\pi^2}\ln\dfrac{M_R}{M_Z} \cr
& = & \dfrac{1}{g_L^2(M_U)} 
+ \dfrac{a_L^\mathrm{IV}}{8\pi^2}\ln\dfrac{M_U}{M_D}
+ \dfrac{a_L^\mathrm{III}}{8\pi^2}\ln\dfrac{M_D}{M_C}
+ \dfrac{a_L^\mathrm{II}}{8\pi^2}\ln\dfrac{M_C}{M_R}
+ \dfrac{a_2^\mathrm{I}}{8\pi^2}\ln\dfrac{M_R}{M_Z} 
\;,
\cr
\dfrac{1}{e^2(M_Z)}
& = & \dfrac{1}{g_2^2(M_Z)} + \dfrac{1}{g_1^2(M_Z)} \cr
& = & \dfrac{1}{g_2^2(M_R)} + \dfrac{1}{g_1^2(M_R)} 
+ \dfrac{1}{8\pi^2}\left(a_1 + a_2\right)^\mathrm{I}\ln\dfrac{M_R}{M_Z} \cr
& = & \dfrac{1}{g_L^2(M_R)} + \dfrac{1}{g_R^2(M_R)}+ \dfrac{1}{g_{BL}^2(M_R)}
+ \dfrac{1}{8\pi^2}\left(a_1 + a_2\right)^\mathrm{I}\ln\dfrac{M_R}{M_Z} \cr
& = & \dfrac{1}{g_L^2(M_C)} + \dfrac{1}{g_R^2(M_C)}+ \dfrac{1}{g_{BL}^2(M_C)} 
+ \dfrac{1}{8\pi^2}\left(a_L + a_R + a_{BL}\right)^\mathrm{II}\ln\dfrac{M_C}{M_R} 
\cr
& & 
+ \dfrac{1}{8\pi^2}\left(a_1 + a_2\right)^\mathrm{I}\ln\dfrac{M_R}{M_Z} 
\cr
& = & \dfrac{1}{g_L^2(M_C)} + \dfrac{1}{g_R^2(M_C)}+ \dfrac{2}{3}\dfrac{1}{g_{4}^2(M_C)} 
\cr
& & 
+ \dfrac{1}{8\pi^2}\left(a_L + a_R + a_{BL}\right)^\mathrm{II}\ln\dfrac{M_C}{M_R} 
+ \dfrac{1}{8\pi^2}\left(a_1 + a_2\right)^\mathrm{I}\ln\dfrac{M_R}{M_Z} 
\cr
& = & \dfrac{1}{g_L^2(M_D)} + \dfrac{1}{g_R^2(M_D)}+ \dfrac{2}{3}\dfrac{1}{g_{4}^2(M_D)} 
+ \dfrac{1}{8\pi^2}\left(a_L + a_R + \dfrac{2}{3}\,a_4\right)^\mathrm{III}\ln\dfrac{M_D}{M_C}
\cr
& & 
+ \dfrac{1}{8\pi^2}\left(a_L + a_R + a_{BL}\right)^\mathrm{II}\ln\dfrac{M_C}{M_R} 
+ \dfrac{1}{8\pi^2}\left(a_1 + a_2\right)^\mathrm{I}\ln\dfrac{M_R}{M_Z} 
\cr
& = & \dfrac{1}{g_L^2(M_U)} + \dfrac{1}{g_R^2(M_U)}+ \dfrac{2}{3}\dfrac{1}{g_{4}^2(M_U)} 
\cr
& & 
+ \dfrac{1}{8\pi^2}\left(a_L + a_R + \dfrac{2}{3}\,a_4\right)^\mathrm{IV}\ln\dfrac{M_U}{M_D}
+ \dfrac{1}{8\pi^2}\left(a_L + a_R + \dfrac{2}{3}\,a_4\right)^\mathrm{III}\ln\dfrac{M_D}{M_C}
\cr
& & 
+ \dfrac{1}{8\pi^2}\left(a_L + a_R + a_{BL}\right)^\mathrm{II}\ln\dfrac{M_C}{M_R} 
+ \dfrac{1}{8\pi^2}\left(a_1 + a_2\right)^\mathrm{I}\ln\dfrac{M_R}{M_Z} 
\;,
\cr
\dfrac{1}{g_3^2(M_Z)}
& = & \dfrac{1}{g_3^2(M_R)} + \dfrac{a_3^\mathrm{I}}{8\pi^2}\ln\dfrac{M_R}{M_Z} \cr
& = & \dfrac{1}{g_3^2(M_C)} + \dfrac{a_3^\mathrm{II}}{8\pi^2}\ln\dfrac{M_R}{M_Z} 
+ \dfrac{a_3^\mathrm{I}}{8\pi^2}\ln\dfrac{M_R}{M_Z} \cr
& = & \dfrac{1}{g_4^2(M_C)} 
+ \dfrac{a_3^\mathrm{II}}{8\pi^2}\ln\dfrac{M_C}{M_R} 
+ \dfrac{a_3^\mathrm{I}}{8\pi^2}\ln\dfrac{M_R}{M_Z} \cr
& = & \dfrac{1}{g_4^2(M_D)} 
+ \dfrac{a_4^\mathrm{III}}{8\pi^2}\ln\dfrac{M_D}{M_C} 
+ \dfrac{a_3^\mathrm{II}}{8\pi^2}\ln\dfrac{M_C}{M_R} 
+ \dfrac{a_3^\mathrm{I}}{8\pi^2}\ln\dfrac{M_R}{M_Z} \cr
& = & \dfrac{1}{g_4^2(M_U)} 
+ \dfrac{a_4^\mathrm{IV}}{8\pi^2}\ln\dfrac{M_U}{M_D} 
+ \dfrac{a_4^\mathrm{III}}{8\pi^2}\ln\dfrac{M_D}{M_C} 
+ \dfrac{a_3^\mathrm{II}}{8\pi^2}\ln\dfrac{M_C}{M_R} 
+ \dfrac{a_3^\mathrm{I}}{8\pi^2}\ln\dfrac{M_R}{M_Z} 
\;,
\cr
\dfrac{1}{g_R^2(M_R)}
& = & \dfrac{1}{g_R^2(M_C)} + \dfrac{a_R^\mathrm{II}}{8\pi^2}\ln\dfrac{M_C}{M_R} \cr
& = & \dfrac{1}{g_R^2(M_D)} 
+ \dfrac{a_R^\mathrm{III}}{8\pi^2}\ln\dfrac{M_D}{M_C}
+ \dfrac{a_R^\mathrm{II}}{8\pi^2}\ln\dfrac{M_C}{M_R} \cr
& = & \dfrac{1}{g_R^2(M_U)} 
+ \dfrac{a_R^\mathrm{IV}}{8\pi^2}\ln\dfrac{M_U}{M_D}
+ \dfrac{a_R^\mathrm{III}}{8\pi^2}\ln\dfrac{M_D}{M_C}
+ \dfrac{a_R^\mathrm{II}}{8\pi^2}\ln\dfrac{M_C}{M_R}
\;.
\end{eqnarray}
If we impose the condition
\begin{equation}
g_L(M_U)\;=\;g_R(M_U)\;=\;g_4(M_U)\;\equiv\; g_U\;,
\end{equation}
then it is straightforward to show that
\begin{eqnarray}
\lefteqn{
2\pi\left[\dfrac{3-8\sin^2\theta_W(M_Z)}{\alpha(M_Z)}\right]
}\cr
& = & 8\pi^2\left[\dfrac{3}{e^2(M_Z)} - \dfrac{8}{g_2^2(M_Z)}\right] \cr
& = & 
\Biggl[
 \left(-5a_L+3a_R+2a_4\right)^\mathrm{IV}\ln\dfrac{M_U}{M_D}
+\left(-5a_L+3a_R+2a_4\right)^\mathrm{III}\ln\dfrac{M_D}{M_C}
\cr
& &
+\left(-5a_L + 3a_R + 3a_{BL}\right)^\mathrm{II}\ln\dfrac{M_C}{M_R}
+\left(3a_1 -5a_2\right)^\mathrm{I}\ln\dfrac{M_R}{M_Z}
\Biggr]
\;,
\cr
\lefteqn{
2\pi\left[\dfrac{3}{\alpha(M_Z)} - \dfrac{8}{\alpha_s(M_Z)}\right]
}\cr
& = & 8\pi^2\left[\dfrac{3}{e^2(M_Z)} - \dfrac{8}{g_3^2(M_Z)}\right] \cr
& = & 
\Biggl[
 \left(3a_L+3a_R-6a_4\right)^\mathrm{IV}\ln\dfrac{M_U}{M_D}
+\left(3a_L+3a_R-6a_4\right)^\mathrm{III}\ln\dfrac{M_D}{M_C}
\cr
& &
+\left(3a_L + 3a_R + 3a_{BL} - 8a_3\right)^\mathrm{II}\ln\dfrac{M_C}{M_R}
+\left(3a_1 + 3a_2 - 8a_3\right)^\mathrm{I}\ln\dfrac{M_R}{M_Z}
\Biggr]
\;,
\cr
\lefteqn{
2\pi\left[\dfrac{4\pi}{g_R^2(M_R)}-\dfrac{\sin^2\theta_W(M_Z)}{\alpha(M_Z)}\right]
} \cr
& = & 8\pi^2\left[\dfrac{1}{g_R^2(M_R)}-\dfrac{1}{g_2^2(M_Z)}\right] 
\cr
& = & 8\pi^2\left[\dfrac{1}{g_R^2(M_R)}-\dfrac{1}{g_L^2(M_R)}\right]
+ 8\pi^2\left[\dfrac{1}{g_2^2(M_R)}-\dfrac{1}{g_2^2(M_Z)}\right]
\cr
& = & 
\left[
 \left(a_R - a_L\right)^\mathrm{III}\ln\dfrac{M_D}{M_C}
+\left(a_R - a_L\right)^\mathrm{II}\ln\dfrac{M_C}{M_R}
-a_2^\mathrm{I}\ln\dfrac{M_R}{M_Z}
\right]
\;,
\cr
\dfrac{8\pi^2}{g_U^2}
& = & \dfrac{3}{8}
\Biggl[
\dfrac{2\pi}{\alpha(M_Z)}
-\Biggl\{
  \left(a_L + a_R + \dfrac{2}{3}\,a_4\right)^\mathrm{IV}\ln\dfrac{M_U}{M_D}
+ \left(a_L + a_R + \dfrac{2}{3}\,a_4\right)^\mathrm{III}\ln\dfrac{M_D}{M_C}
\cr
& & \qquad
+ \left(a_L + a_R + a_{BL}\right)^\mathrm{II}\ln\dfrac{M_C}{M_R} 
+ \left(a_1 + a_2\right)^\mathrm{I}\ln\dfrac{M_R}{M_Z} 
\Biggr\}
\Biggr]
\cr
& = & \dfrac{2\pi}{\alpha_s(M_Z)}
-\left(
  a_4^\mathrm{IV}\;\ln\dfrac{M_U}{M_D} 
+ a_4^\mathrm{III}\;\ln\dfrac{M_D}{M_C} 
+ a_3^\mathrm{II}\;\ln\dfrac{M_C}{M_R} 
+ a_3^\mathrm{I}\;\ln\dfrac{M_R}{M_Z} 
\right)
\;.
\end{eqnarray}
Note that $a_L^\mathrm{IV}=a_R^\mathrm{IV}$ since parity is not broken in energy interval IV.

If instead, we impose the conditions
\begin{equation}
g_L(M_D) \;=\; g_R(M_D) \;\equiv\; g_{LR}\;,\qquad
g_4(M_D) \;\equiv\; g_{4D}\;,
\end{equation}
where $g_{LR}$ and $g_{4D}$ are not necessarily equal, then
the relations will be
\begin{eqnarray}
\lefteqn{
2\pi
\left[
\dfrac{3-6\sin^2\theta_W(M_Z)}{\alpha(M_Z)}-\dfrac{2}{\alpha_s(M_Z)}
\right]
}
\cr
& = & 8\pi^2
\left[
\dfrac{3}{e^2(M_Z)} - \dfrac{6}{g_2^2(M_Z)} - \dfrac{2}{g_3^2(M_Z)}
\right]
\cr
& = & 
\Biggl[
 \left(-3a_L+3a_R\right)^\mathrm{III}\ln\dfrac{M_D}{M_C}
+\left(-3a_L+3a_R+3a_{BL}-2a_3\right)^\mathrm{II}\ln\dfrac{M_C}{M_R}
\cr & & \quad
+\left(3a_1-3a_2-2a_3\right)^\mathrm{I}\ln\dfrac{M_R}{M_Z}
\Biggr]
\;,
\cr
\lefteqn{
2\pi\left[\dfrac{4\pi}{g_R^2(M_R)}-\dfrac{\sin^2\theta_W(M_Z)}{\alpha(M_Z)}
\right]
} \cr
& = & 8\pi^2
\left[\dfrac{1}{g_R^2(M_R)}-\dfrac{1}{g_2^2(M_Z)}\right]
\cr
& = & \left[\dfrac{1}{g_R^2(M_R)}-\dfrac{1}{g_L^2(M_R)}\right]
+ \left[\dfrac{1}{g_2^2(M_R)}-\dfrac{1}{g_2^2(M_Z)}\right]
\cr
& = & \dfrac{1}{8\pi^2}
\left[
 \left(a_R - a_L\right)^\mathrm{III}\ln\dfrac{M_D}{M_C}
+\left(a_R - a_L\right)^\mathrm{II}\ln\dfrac{M_C}{M_R}
-a_2^\mathrm{I}\ln\dfrac{M_R}{M_Z}
\right]
\;,
\cr
\dfrac{8\pi^2}{g_{LR}^2}
& = & \dfrac{\pi}{\alpha(M_Z)}
-\dfrac{1}{3}\dfrac{2\pi}{\alpha_s(M_Z)}
-\dfrac{1}{6}\Biggl[
\left\{
3\left(a_L + a_R \right)
\right\}^\mathrm{III}\;\ln\dfrac{M_D}{M_C}
\cr
& &
+\left\{
3\left(a_L + a_R + a_{BL}\right)
-2a_3
\right\}^\mathrm{II}\;\ln\dfrac{M_C}{M_R}
+\left\{
3\left(a_1+a_2\right)
-2a_3
\right\}^\mathrm{I}\;\ln\dfrac{M_R}{M_Z}
\Biggr]
\;,
\cr
\dfrac{8\pi^2}{g_{4D}^2}
& = & \dfrac{2\pi}{\alpha_s(M_Z)}
-\left(
  a_4^\mathrm{III}\;\ln\dfrac{M_D}{M_C} 
+ a_3^\mathrm{II}\;\ln\dfrac{M_C}{M_R} 
+ a_3^\mathrm{I}\;\ln\dfrac{M_R}{M_Z} 
\right)
\;.
\end{eqnarray}
%


\end{document}